\documentclass[useAMS,usenatbib,usegraphicx,letterpaper]{mn2e}
\usepackage{times,amssymb,hyperref,subfigure,epsfig,aas_macros}

\newcommand{\kms}{km~s$^{-1}$}

\begin{document}

\title[The HD~166191 transition disc]{Evolution from protoplanetary to debris discs: \\
  The transition disc around HD~166191}

\author[G. M. Kennedy et al.]{G. M. Kennedy\thanks{Email:
    \href{mailto:gkennedy@ast.cam.ac.uk}{gkennedy@ast.cam.ac.uk}}$^1$, S. J. Murphy$^2$,
  C. M. Lisse$^3$, F. M\'enard$^4$, M. L. Sitko$^5$, M. C. Wyatt$^1$, \newauthor
  D. D. R. Bayliss$^6$, F. E. DeMeo$^7$, K. B. Crawford$^8$, D. L. Kim$^9$, R. J. Rudy$^8$,
  \newauthor R. W. Russell$^9$, B. Sibthorpe$^{10}$, M. A. Skinner$^{11}$, G. Zhou$^{6}$ \\
  $^1$ Institute of Astronomy, University of Cambridge, Madingley Road, Cambridge CB3
  0HA, UK \\
  $^2$ Astronomisches Rechen-Institut, Zentrum f\"ur Astronomie der Universit\"at
  Heidelberg, D-69120 Heidelberg, Germany \\
  $^3$ JHU-APL, 11100 Johns Hopkins Road, Laurel, MD 20723, USA \\
  $^4$ UMI-FCA, CNRS/INSU, France (UMI 3386), and Dept. de Astronom\'{\i}a, Universidad de Chile, Santiago, Chile \\
  $^5$ Department of Physics, University of Cincinnati, Cincinnati, OH 45221-0011, USA \\
  $^6$ Research School of Astronomy and Astrophysics, The Australian National University, Mount Stromlo Observatory, Cotter Road, Weston Creek, ACT 2611, Australia \\
  $^7$ Massachusetts Institute of Technology, Department of Earth, Atmospheric and Planetary Sciences, 77 Massachusetts Avenue, Cambridge, MA 02139, USA \\
  $^8$ The Aerospace Corporation, 2310 E. El Segundo Blvd. El Segundo, CA, USA \\
  $^9$ The Aerospace Corporation, Mail Stop: M2-266, P. O. Box 92957, Los Angeles, CA
  90009-2957 \\
  $^{10}$ SRON Netherlands Institute for Space Research, NL-9747 AD Groningen, The
  Netherlands \\
  $^{11}$ The Boeing Company, 535 Lipoa Pkwy, Kihei, HI 96753, USA \\}
\maketitle

\begin{abstract}
  HD~166191 has been identified by several studies as hosting a rare and extremely bright
  warm debris disc with an additional outer cool disc component. However, an alternative
  interpretation is that the star hosts a disc that is currently in transition between a
  full gas disc and a largely gas-free debris disc. With the help of new optical to
  mid-IR spectra and \emph{Herschel} imaging, we argue that the latter interpretation is
  supported in several ways: i) we show that HD~166191 is co-moving with the
  $\sim$4~Myr-old Herbig Ae star HD~163296, suggesting that the two have the same age,
  ii) the disc spectrum of HD~166191 is well matched by a standard radiative transfer
  model of a gaseous protoplanetary disc with an inner hole, and iii) the HD~166191
  mid-IR silicate feature is more consistent with similarly primordial objects. We note
  some potential issues with the debris disc interpretation that should be considered for
  such extreme objects, whose lifetime at the current brightness is much shorter than the
  stellar age, or in the case of the outer component requires a mass comparable to the
  solid component of the Solar nebula. These aspects individually and collectively argue
  that HD~166191 is a 4-5~Myr old star that hosts a gaseous transition disc. Though it
  does not argue in favour of either scenario, we find strong evidence for 3-5~$\mu$m
  disc variability. We place HD~166191 in context with discs at different evolutionary
  stages, showing that it is a potentially important object for understanding the
  protoplanetary to debris disc transition.
\end{abstract}

\begin{keywords}
  protoplanetary discs --- circumstellar matter --- planets and satellites: formation ---
  stars: individual: HD~166191 --- stars: individual: HD~163296
\end{keywords}

\section{Introduction}\label{s:intro}

As nurseries in which planets form, protoplanetary discs are all-important in setting the
boundary conditions for planet formation. One of their most important characteristics is
the disc lifetime because once the gaseous component has dispersed, giant planet
formation is halted. Therefore, the lifetime of protoplanetary discs and how they
disperse has been the focus of many studies
\citep[e.g.][]{1995Natur.373..494Z,2001MNRAS.328..485C,2001ApJ...553L.153H,2004ApJ...612..496M,2004ApJ...611..360A,2012ApJ...756..133C}
and sets hard constraints for models of giant planet formation.

The nature of the transition from a gaseous protoplanetary disc to a relatively gas-free
debris disc is poorly understood, particularly the impact on the formation and evolution
of planets and planetesimals. Observations at a range of infrared (IR) wavelengths are
therefore vital to build up possible evolutionary sequences. Near/mid-IR observations
probe dispersal and the creation of inner holes in gas-rich ``transition'' discs
\citep{1990AJ.....99.1187S,2010ApJ...708.1107M,2012ApJ...747..103E}, and are also
sensitive to warm dust created during the later stages of terrestrial planet formation
\citep[e.g.][]{2010ApJ...717L..57M,2012MNRAS.tmp.3462J}. Far-IR observations are most
sensitive to changes in the outer regions of protoplanetary discs as dust grains settle
and grow, and to the same $\sim$100AU scales at which debris discs are typically observed
around main-sequence stars.

For some evolved objects it is not clear whether the stars are still dispersing their gas
and are in the ``transition'' phase, or if the disc is sufficiently gas-free that it can
be considered a true debris disc. Perhaps the best example is the extremely bright disc
around HD~98800B \citep[$L_{\rm disc}/L_\star \approx 2\%$,][]{1993ApJ...406L..25Z},
which has been treated as either by different authors and illustrates that classification
is not straightforward \citep{2007ApJ...664.1176F,2007ApJ...658..569W}. The recent
discovery of H$_2$ emission, a signature of gas accretion, seems to favour the transition
disc interpretation \citep{2012ApJ...744..121Y}. Of course, there will be a continuum of
objects between relatively gas-rich and gas-free discs, and no disc could ever be
considered truly devoid of gas, but in order to create a useful evolutionary sequence it
is important to interpret individual objects carefully. Doing so will, for example,
ensure that unlikely scenarios are not invoked where simpler ones will suffice, and that
observational sequences of objects are as close to reality as possible.

Here, we study an object that we suggest lies at the late end of the transition to a
gas-free debris disc. HD~166191 has been known to have an IR excess above the expected
photospheric level since it was observed with IRAS and MSX
\citep{1992A&AS...96..625O,2005MNRAS.363.1111C}. The possibility of confusion of the IRAS
emission with another nearby source, which is visible to the South-West in higher
resolution images, meant that the excess was only detected unambiguously from
8-20~$\mu$m. More recently, this system was suggested to have a mid-IR excess with both
WISE \citep{2013MNRAS.433.2334K} and AKARI \citep{2013A&A...550A..45F}, which was
confirmed with high resolution mid-IR imaging \citep{2013arXiv1308.0405S}. While these
studies went in search of warm terrestrial-zone debris discs, only
\citet{2013arXiv1308.0405S} used the Multi-band Imaging Photometer for \emph{Spitzer}
\citep[MIPS,][]{2004ApJS..154....1W,2004ApJS..154...25R} to show that the far-IR emission
is probably associated with the star, and therefore that the disc emission is significant
over a wide range of wavelengths.

We previously argued that this star hosts a gas-rich disc based on the breadth of the
disc spectrum and its brightness relative to the star \citep{2013MNRAS.433.2334K}, but
the other studies \citep{2013A&A...550A..45F,2013arXiv1308.0405S} have interpreted it as
hosting a debris disc with two components (i.e. analogous to the Asteroid and
Edgeworth-Kuiper belts seen in our Solar System). Here we use several lines of evidence
to argue that HD~166191 does indeed host a gaseous protoplanetary disc. After discussing
the observations in \S \ref{s:obs}, we show in \S \ref{s:age} that HD~166191 has a common
space motion with the $\sim$4~Myr-old Herbig Ae star HD~163296 suggesting that HD~166191
has a similar age. In \S \ref{s:rt} we highlight some potential issues with the debris
disc interpretation and show that the HD~166191 disc emission is well modelled by a
relatively normal transition disc, as might be expected for such a young star. Finally,
we compare the HD~166191 disc to others, taking a close look at the 10~$\mu$m silicate
feature and the overall disc spectrum, and discuss the implications of our work in \S
\ref{s:disc}.

\section{Observations}\label{s:obs}

\begin{figure}
  \begin{center}
    \hspace{-0.5cm} \includegraphics[width=0.5\textwidth]{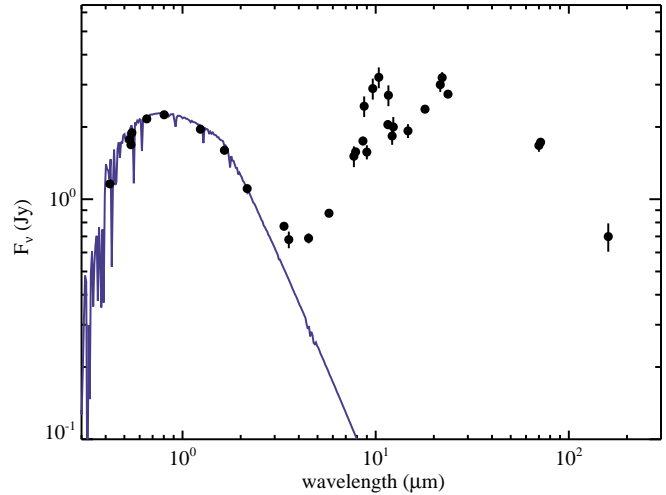}
    \caption{Photometry (dots) and the 6170K PHOENIX AMES-Cond photosphere model (blue
      curve) for HD~166191.}\label{fig:phot}
  \end{center}
\end{figure}

HD~166191 is an 8.4th magnitude star that lies 119pc from the Sun
\citep{2007A&A...474..653V} in the direction of the Galactic center. The Michigan
spectral type is F4V, though using new data we suggest below that the star is actually
closer to a late-F or early G-type, similar to the F8 spectral type found by
\citet{2013arXiv1308.0405S}. The star was detected by both Hipparcos and 2MASS, which we
combine with $BVRI$ photometry from \citet{1984ApJS...55..389M} to fit a stellar
photosphere model. We fitted PHOENIX AMES-Cond models \citep{2005ESASP.576..565B} using
least-squares minimisation, finding an effective temperature of $6170 \pm 50$K. A wealth
of photometry exists for HD~166191, and is tabulated by \citet{2013arXiv1308.0405S} and
shown in Figure \ref{fig:phot}, including new \emph{Herschel} observations described
below. We do not use the WISE $4.6\mu$m (W2) measurement, as this is known to be
overestimated for bright
sources.\footnote{http://wise2.ipac.caltech.edu/docs/release/allsky/expsup/sec6\_3c.html}

\subsection{New observations}\label{ss:newobs}

To ensure that the far-IR excess was not confused in the MIPS 70~$\mu$m observations and
to constrain the disc size we obtained 70 and 160~$\mu$m observations with PACS instrument
\citep{2010A&A...518L...2P} on the Herschel Space Observatory \citep{2010A&A...518L...1P}
in March 2013. In addition, we obtained ground-based optical, near, and mid-IR spectra
with the goal of better characterising the star and the near/mid-IR excess.

\subsubsection{SSO 2.3m WiFeS}\label{ss:sso}

We observed HD~166191 with the Wide Field Spectrograph
\citep[WiFeS,][]{2007Ap&SS.310..255D} on the ANU 2.3m telescope at Siding Spring
Observatory. Flux calibrations are performed according to \citet{1999PASP..111.1426B}
using spectrophotometric standard stars from \citet{1992PASP..104..533H} and
\citet{1999PASP..111.1426B}. A full description of the instrument configurations and data
reduction procedure can be found in \citet{2013AJ....145....5P}. We use the
$\lambda/\Delta\lambda = R = 3000$ blue arm spectrum to estimate the stellar effective
temperature by comparison with model atmospheres. The method compares the spectrum with
synthetic spectral models, with extra weight given to regions that are sensitive to
surface gravity, and is described in detail by \citet{2013arXiv1306.0624B}. Assuming
Solar metallicity we obtain an effective temperature of $5965 \pm 100$K, in reasonable
agreement with the value drived from broadband photometry above. The gravity is found to
be $\log g = 3.4 \pm 0.25$, suggesting that HD~166191 has not yet reached the
main-sequence (see \S \ref{s:age}).

In the WiFeS $R = 7000$ red arm spectrum Li\,\textsc{i} absorption at 6708\AA~(an
indicator of youth in Solar type stars, see \S \ref{s:age}) is detected with an
equivalent width of $150 \pm 20$m\AA. From these observations we also obtained a radial
velocity of $-10.1 \pm 1$km s$^{-1}$, in agreement with $-8.1 \pm 1.3$km s$^{-1}$
reported by \citet{2013arXiv1308.0405S}. Hydrogen $\alpha$ and $\beta$, and Sodium
doublet absorption is seen, so any gas accretion, which would result in emission
\citep[e.g.][]{2001ApJ...550..944M}, is below a detectable level
\citep[$\lesssim$10$^{-10}$ to $10^{-11}M_\odot {\rm yr}^{-1}$,
e.g.][]{2003ApJ...592..266M,2006AJ....132.2135S}.

\subsubsection{IRTF/SPeX}\label{sss:spex}

\begin{figure}
  \begin{center}
    \hspace{-0.5cm} \includegraphics[width=0.5\textwidth]{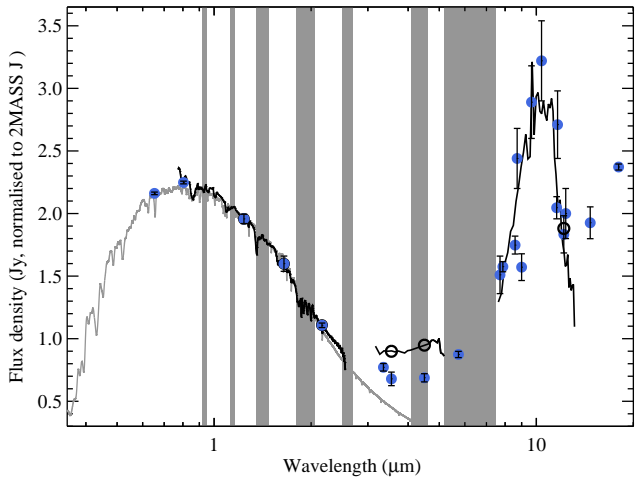}
    \caption{IRTF SpeX and BASS spectra, and a 6170K model atmosphere convolved to
      approximately the same resolution. Filled blue dots show observed photometry and
      open circles show synthetic photometry of the spectra for comparison, where
      possible (MSX 12$\mu$m, and IRAC 3.5 and 4.5~$\mu$m). The SPeX spectra is
      normalised to 2MASS J. Regions of large telluric absorption where the spectrum may
      be adversely effected are blocked out in grey. A small IR excess may be visible
      beyond about 2.1~$\mu$m in the SPeX spectrum, and clearly visible in the BASS
      spectrum. The difference between the IRAC 3.5 and 4.5~$\mu$m photometry and the
      equivalent spectrophotometry from BASS are significantly different, and indicate
      probable variability in the excess spectrum.}\label{fig:spex}
  \end{center}
\end{figure}

To further characterise the near-IR part of the excess, near-infrared spectral
measurements from 0.8 to 2.5~$\mu$m were obtained using the SpeX instrument on the 3 m
NASA Infrared Telescope Facility (IRTF) located on Mauna Kea, Hawaii. It was operated in
single prism mode (R=250). Frames were taken so that the object was alternated between
two different positions (usually noted as the `A' and `B' positions) on a $0.8 \times 15$
arcsec slit aligned North-South. Two A0V stars were observed for telluric corrections,
HD~168707 and HD~170364. Argon lamps were used for wavelength calibration. 24 images were
taken with 1 second exposures and 10 coadds and 16 images were taken with 0.5 second
exposures and 10 coadds. The airmass ranged from 1.41 to 1.47 during the observations.

The spectrum is shown in Figure \ref{fig:spex}, along with a 6170K model atmosphere
(i.e. the same stellar model used in Figure \ref{fig:phot}). At 2MASS $K_s$ and beyond
the possible onset of the IR excess is seen. An excess at these wavelengths is expected
given that an excess is confidently detected beyond 3~$\mu$m.

By comparing the spectrum with the \citet{2009ApJS..185..289R} spectral atlas the
spectral type appears to be late-F or early G-type. The Paschen $\delta$, $\gamma$, and
$\beta$, and Brackett $\gamma$ lines are clearly visible in the spectrum, and the atlas
shows that these are present in F9V stars but less clear by G2V. Based on these lines and
the effective temperatures derived above, we assign HD~166191 a G0V spectral type, with
an uncertainty of one-two subtypes.

\subsubsection{AEOS/BASS}

Observations were obtained using the Aerospace Corporation's Broadband Array Spectrograph
System \citep[BASS,][]{1990SPIE.1235..171H} instrument on the 3.67 m Advanced Electro
Optical System (AEOS) telescope on Haleakala, Hawaii. This instrument uses a cold beam
splitter to separate the light into two separate wavelength regimes. The short-wavelength
beam includes light from 2.9 to 6~$\mu$m, while the long-wavelength beam covers
6-13.5~$\mu$m. Each beam is dispersed onto a 58 element blocked impurity band linear
array, thus allowing for simultaneous coverage of the spectrum from 2.9-13.5~$\mu$m. The
spectral resolution is wavelength dependent, ranging from about $R = 30$ to 125 over each
of the two wavelength regions. The simultaneous coverage of the entire 3-13~$\mu$m region
that BASS provides has played a critical role in the investigations of the clearing of
the inner discs \citep{2002ApJ...568.1008C,2004ApJ...614L.133B,2008ApJ...678.1070S}.

The BASS data were calibrated using $\alpha$ Lyrae and $\beta$ Peg and cleaned of
residual telluric (CO$_2$ and H$_2$O) contamination, but that of O$_3$ at 9.7 $\mu$m was
left in. The spectrum is shown in Figure \ref{fig:spex} and shows a clear silicate
feature around 10~$\mu$m and is in good agreement with the T-ReCS photometry presented in
\citet{2013arXiv1308.0405S}. In addition, the absolute calibration of BASS spectra is
known to be better than about 5\% \citep{basscalib}, meaning that we can test for
differences between the BASS spectrum and the older IRAC data (2006) at 3.5 to 4.5~$\mu$m
(also shown in Figure \ref{fig:spex}). The BASS/IRAC flux ratios in these bands are 1.32
and 1.38, with significances of 3.1 and 4.5$\sigma$, indicative of significant
variability in the excess spectrum. \citet{2013arXiv1308.0405S} note the possibility of
variability based on the difference between IRAC and WISE photometry at $4.6\mu$m, but
WISE is known to overestimate fluxes for bright sources in this band (see first
footnote).

% print,0.899/0.6789,0.948/0.6875

\subsubsection{Herschel}

Following identification of this object in \citet{2013MNRAS.433.2334K}, we obtained a
Director's time observation of HD~166191 using the PACS instrument onboard
\emph{Herschel} (ObsIDs: 1342267772/3). We used a slightly non-standard mini-scan map,
designed to be wide enough to cover both HD~166191 and the nearby source that
contaminates the IRAS photometry. The flux density of both objects was measured by
fitting a model beam (an observation of $\gamma$ Dra) at each location. We found fluxes
of $1.7 \pm 0.1$Jy and $0.75 \pm 0.1$Jy at 70 and 160~$\mu$m for HD~166191, consistent
with the \emph{Spitzer} photometry at 70~$\mu$m presented by
\citet{2013arXiv1308.0405S}. This observation further confirms that HD~166191 has a
significant far-IR excess (see Figure \ref{fig:phot}) and because the emission is
point-like (with beam size $\sim$5\arcsec) an approximate upper limit on the source
diameter is $\sim$5\arcsec, or about 600AU.

\subsection{Summary of observations}

Based on previous data and the new observations described above we assign the star a
spectral type of G0V. An IR excess is clearly visible beyond 3~$\mu$m, with a clear
10~$\mu$m silicate feature. A significant discrepancy between the IRAC and BASS
observations at 3-5~$\mu$m suggests that the disc emission is variable, which is not
unusual for transition discs
\citep{2009ApJ...706L.168B,2009ApJ...704L..15M,2011ApJ...728...49E}. The very large disc
luminosity ($L_{\rm disc}/L_\star \approx 10$\%), the fact that the excess spectrum
extends over a wide range of wavelengths (Fig \ref{fig:phot}), and the probable
variability suggests at first glance that the emission is probably from a young
($\lesssim$10~Myr old) star that hosts a gas-rich protoplanetary disc. However, it is also
known that some debris disc brightnesses vary
\citep[e.g.][]{2012ApJ...751L..17M,2012Natur.487...74M,2013MNRAS.433.2334K}, and an
alternative interpretation could be that HD~166191 hosts an unusually bright debris disc
with both warm and cool components \citep{2013A&A...550A..45F,2013arXiv1308.0405S}, in
essence an extreme analogue of the two-component debris disc around $\eta$ Corvi
\citep{2005ApJ...620..492W,2009A&A...503..265S}. The most promising way to discern
between these possibilities and to find a self-consistent explanation is to determine the
age of HD~166191.

\section{The age of HD~166191}\label{s:age}

There are several techniques available to infer the ages of young stars
\citep[e.g.][]{2004ARA&A..42..685Z,2010ARA&A..48..581S}. In this section we attempt to
refine the age of HD~166191 using the empirical methods of lithium depletion and X-ray
activity, combined with kinematic considerations and (model-dependent) isochrone
placement.

Because it is efficiently destroyed in stellar interiors, photospheric lithium abundance
can serve as a mass-dependent clock over pre-main sequence ages
\citep{2004ARA&A..42..685Z,2013MNRAS.435.1325M}. Our Li\,\textsc{i} $\lambda$6708
equivalent width, $150\pm 20$~m\AA, agrees with the high-resolution value reported by
\citet{2013arXiv1308.0405S} ($120\pm5$~m\AA). At the spectral type of HD~166191 this
value is consistent with an age less than the Pleiades
\citep[$\lesssim$100~Myr;][]{2004ARA&A..42..685Z}. The poor age constraint of lithium
depletion in Solar-type stars is due to their radiative cores effectively separating the
convective outer layers from the hotter interiors, preventing lithium depleted material
from reaching the photosphere.

X-ray emission is another common diagnostic of stellar youth, a manifestation of the
strong magnetic fields and fast rotation rates observed in young stars
\citep[e.g.][]{1999ARA&A..37..363F}. HD~166191 was not detected in the \emph{ROSAT}
All-Sky X-ray Survey (RASS).  The RASS 0.1--2.4 eV limiting flux \citep[$2\times10^{-13}$
erg cm$^{-2}$ s$^{-1}$;][]{1995ApJ...450..392S} at 119~pc implies an X-ray to bolometric
luminosity ratio of $\log(L_{X}/L_{\rm bol})\lesssim-4.8$. \citet{2013arXiv1308.0405S}
report a detection in the XMM-Newton Serendipitous Source Catalog
\citep{2009A&A...493..339W}, which yields a 0.2--12 eV luminosity of $2.4\times10^{29}$
erg s$^{-1}$ and $\log(L_{X}/L_{\rm bol})\approx -4.9$. Ignoring the subtleties of
comparing results from different X-ray observatories and energy bands, we compare these
values to the results of \citet{2005ApJS..160..390P}, who examined the mass-dependence of
X-ray emission in young stars in Orion as a function of age. Their distributions shows
that while HD~166191 is one of the X-ray faintest 1-$2M_\odot$ stars at 1-10~Myr ages, the
X-ray luminosity also varies by about 2 orders of magnitude at a given age. Comparison
with the X-ray luminosity distribution in \citet{2011MNRAS.412...13O} leads to the same
conclusion. Therefore, though it is indeed X-ray faint as noted by
\citet{2013arXiv1308.0405S}, we do not consider this faintness to provide a significantly
stronger age constraint than the lithium absorption.

\begin{figure*}
  \begin{center}
    \hspace{-0.5cm} \includegraphics[width=1\textwidth]{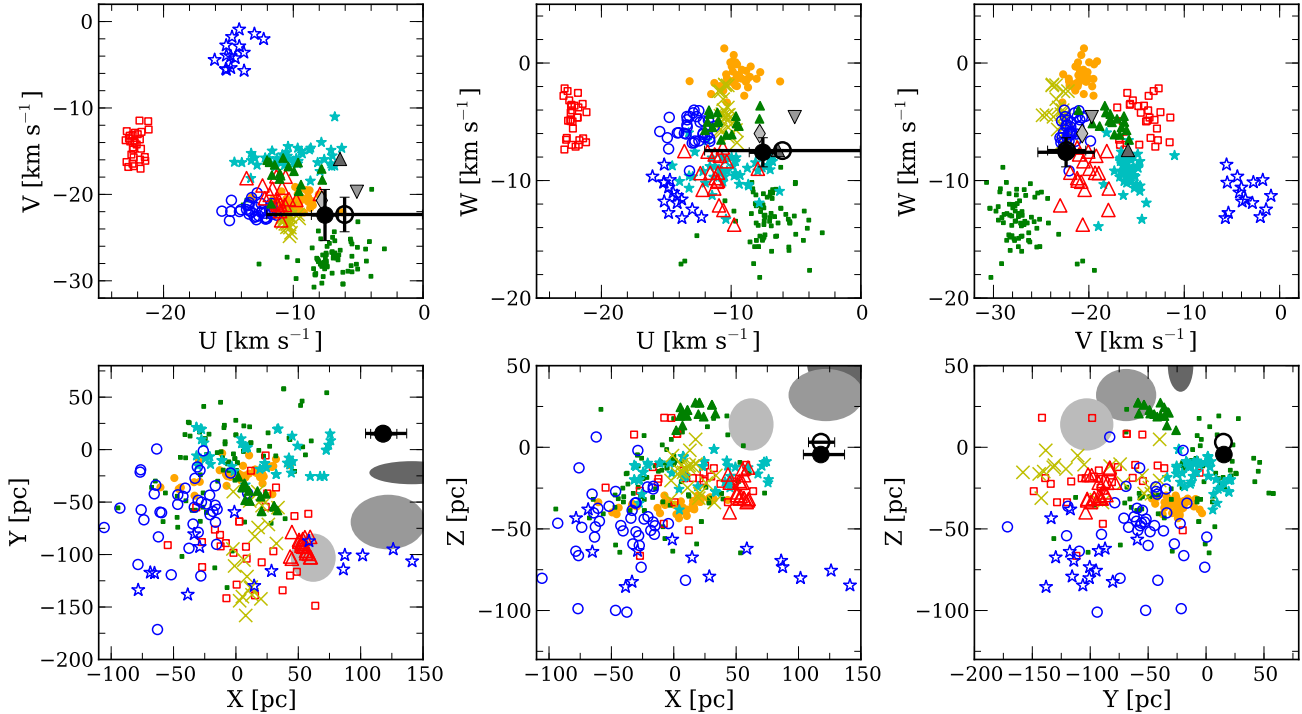}
    \caption{Heliocentric velocity (top) and position (bottom) of HD~166191 (black dot,
      with errors), compared to members of nearby associations and moving groups from
      \citet{2008hsf2.book..757T}: Tuc-Hor (orange circles), AB Dor (green squares),
      Argus (red open squares), $\beta$ Pic (cyan stars), Carina (yellow crosses),
      Columba (blue open circles), $\epsilon$ Cha (red triangles), Octans (blue open
      stars) and TW Hya (green filled triangles). Positions \citep[shaded
      ellipses;][]{2000ApJ...544..356M} and velocities \citep{2011ApJ...738..122C} of the
      three Scorpius-Centaurus OB association subgroups are also plotted: Upper Sco
      (darkest, triangle), Upper Cen-Lup (inverted triangle) and Lower Cen-Cru (lightest,
      diamond). The large open circle in the shows the velocity and position of
      HD~163296.}\label{fig:uvwxyz}
  \end{center}
\end{figure*}

Kinematics are commonly used to assign young stars to moving groups or associations of
known age. From \emph{Hipparcos} astrometry \citep{2007A&A...474..653V} and the
2.3-m/WiFeS radial velocity we calculate a heliocentric space motion for HD~166191 of
$(U,V,W)= (-7.6\pm1.0, -22.4\pm3.0, -7.6\pm1.2)$~\kms. This agrees with the velocity
presented by \citet{2013arXiv1308.0405S} from Tycho-2 proper
motions. Figure~\ref{fig:uvwxyz} shows the position and velocity of HD~166191 compared
other nearby young stars. While the star's space motion is consistent with members of
several young stellar associations, including the Scorpius-Centaurus OB association
(Sco-Cen), its relatively large distance (119$_{-14}^{+19}$~pc) and low Galactic latitude
yields a heliocentric position, $(X,Y,Z)=(+118\pm18, +15\pm2, -4\pm1)$~pc, distinct from
other young groups in the solar neighbourhood.

Like \citet{2013arXiv1308.0405S}, we searched for \emph{Hipparcos} entries around
HD~166191 with similar proper motions and parallaxes and found a match with the
well-studied `isolated' Herbig Ae star HD~163296
(119$_{-10}^{+12}$~pc). \citet{2013arXiv1308.0405S} used the XHIP
\citep{2012AstL...38..331A} radial velocity ($-4.0\pm3.3$~\kms) for HD~163296 from
\citet{1930ApJ....72...98M} via \citet{2006AstL...32..759G}. This is the mean velocity of
a single line (Mg\,\textsc{ii} $\lambda$4481) from seven Mt.\,Wilson plates
($\sigma_{RV}=9.4$~\kms). More recently, \citet{1965MNRAS.130..281B} and
\cite{2013MNRAS.429.1001A} reported larger velocities of $-11$~\kms\ and $-9\pm6$~\kms,
respectively. Using the latter value and \emph{Hipparcos} astrometry we calculate a space
motion of $(U,V,W)= (-6.0\pm6.0, -22.3\pm2.1, -7.4\pm0.8)$~\kms, in excellent agreement
with HD~166191.

\begin{figure}
  \begin{center}
    \hspace{-0.5cm} \includegraphics[width=0.5\textwidth]{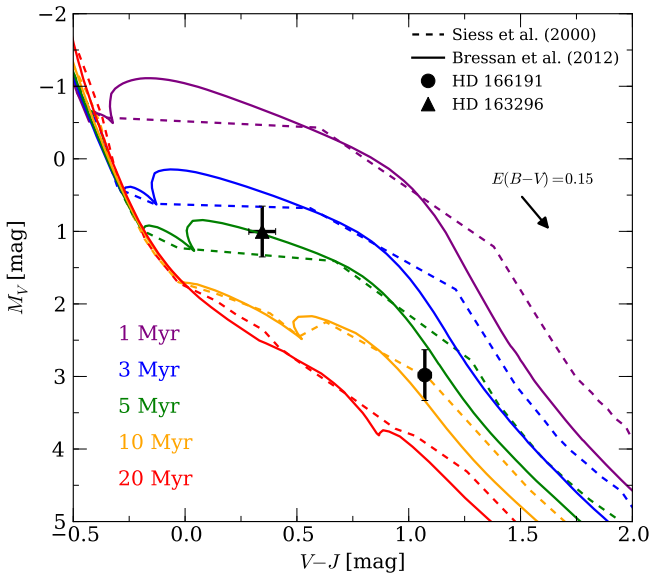} 
    \caption{Colour-magnitude diagram of HD~166191 and HD~163296, with 3--20~Myr
      theoretical isochrones from \citet{2000A&A...358..593S} and
      \citet{2012MNRAS.427..127B} for comparison. For HD~163296, photometry and a
      (circumstellar) extinction of $E(B-V)=0.15$ were adopted from
      \citet{2012A&A...538A..20T}. The Tycho-2 $V$-band magnitude for HD~166191 was
      transformed to the Johnson-Cousins system using the relations of
      \citet{2000PASP..112..961B}. No extinction was adopted for
      HD~166191.}\label{fig:cmd}
  \end{center}
\end{figure}

A colour-magnitude diagram (CMD) of both stars is plotted in Fig.\,\ref{fig:cmd}. If
HD~166191 and HD~163296 are co-eval then they should lie along the same theoretical
isochrone in such a diagram. Compared to PARSEC \citep{2012MNRAS.427..127B} and
\citet{2000A&A...358..593S} isochrones, HD~163296 has an age of 4--5~Myr, consistent with
previous isochronal age estimates from a variety of model grids
\citep{2011MNRAS.410..190T,2012A&A...538A..20T,2013MNRAS.429.1001A}. HD~166191 falls
slightly older, with an age of 5--10~Myr. This age agrees with the (conservative)
$5_{-3}^{+25}$~Myr CMD age reported by \citet{2013arXiv1308.0405S} using the PARSEC
models and 2MASS $K_{s}$ magnitudes. We adopted the $J$-band photometry to minimise the
effects of disc emission observed in both stars at wavelengths greater than 2 $\mu$m (see
\S \ref{sss:spex}). The derived masses of HD~166191 are 1.7 and $1.6M_\odot$ from the
PARSEC and \citet{2000A&A...358..593S} isochrones respectively. Neither star shows signs
of unresolved binarity \citep{2012A&A...538A..20T,2013arXiv1308.0405S}. Given their
proximity (7.5 pc in space), identical distances, congruent space motions and very
similar ages, it is likely HD~166191 and HD~163296 form a co-eval, co-moving pair of
$\sim$4-5~Myr-old stars. HD~163296 was proposed as an outlying member of the
10--20~Myr-old Sco-Cen subgroup Upper Cen-Lup by \citet{2008ApJ...678.1070S}. This
membership is unlikely considering its young age and heliocentric position and velocity
(Fig.\,\ref{fig:uvwxyz}). Instead, we speculate that HD~166191 and HD~163296 may
represent the first members of a yet-to-be-discovered kinematic association or new
Sco-Cen subgroup.
% (c.f. Nguyen et al.\footnote{http://www.mpia-hd.mpg.de/homes/ppvi/posters/1G024.html}).

Having now established the likely youth of HD~166191 using a CMD and by association with
HD~163296, it seems that a less extreme interpretation for the large IR excess may be
that of a gaseous transition disc, in contrast to the debris disc interpretation of
\citet{2013arXiv1308.0405S}.
 
\section{Disc modelling}\label{s:rt}

\subsection{A two-belt debris disc?}\label{ss:2b}

The interpretation favoured by \citet{2013arXiv1308.0405S} is that HD~166191 hosts a warm
terrestrial zone debris disc, the result of collisions between planetesimals or planets
during the late stages of planet formation. However, the total fractional luminosity
$f=L_{\rm disc}/L_\star$ of the excess spectrum is about 10\%, meaning that the dust in a
putative debris disc must capture at least this much of the starlight. If the dust has a
non-negligible albedo or is optically thick along the line of sight the fraction of
starlight captured must be higher. As we show below, because such extreme IR excesses may
require optically thick dust, significantly more than 10\% of the starlight may need to
be captured. An additional complication in the case of HD~166191 is that
\citet{2013arXiv1308.0405S} infer a cool outer component based on the breadth of the
emission spectrum, with a fractional luminosity similar to the inner component. While
such a two-belt structure is of course possible (e.g. the Solar System has two belts),
the extreme brightness of the HD~166191 disc means that the inner belt must capture
$\sim$6\% of the starlight, yet be sufficiently optically thin or misaligned that the
outer belt is still able to capture $\sim$4\%. These constraints set lower limits on the
vertical extent of the inner and outer belts, which can be compared with observations of
edge-on belts around other stars.

\begin{figure}
  \begin{center}
    \hspace{-0.5cm} \includegraphics[width=0.5\textwidth]{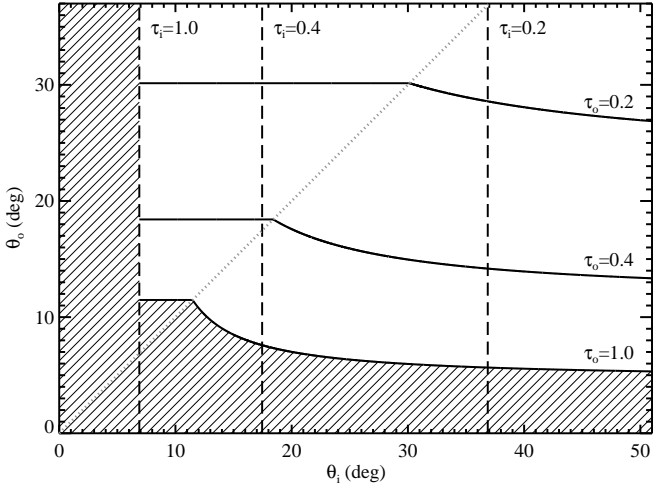}
    \caption{Simple model of allowed opening angles for a coplanar two-belt
      interpretation of HD 166191 with $f_{\rm i}=0.06$ and $f_{\rm o}=0.04$. The dashed
      lines show how the radial optical depth of the inner belt depends on the opening
      angle. The observed fractional luminosity cannot be produced for opening angles in
      the shaded regions (where $\tau_{\rm i,o}>1$). The solid lines show constant radial
      optical depth for the outer belt (eq \ref{eq:outer}). The dotted line shows where
      the belts have equal opening angles.}\label{fig:opang}
  \end{center}
\end{figure}

To make a simple model of the above picture we assume that i) the two belts are coplanar,
ii) have uniform density, iii) the dust in the belts has negligible albedo, iv) the belts
are optically thin from the viewing direction, and v) reemission from either belt does
not heat the star or the other belt. We use parameters $f_{\rm i,o}$ and $\theta_{\rm
  i,o}$ for the fractional luminosity and opening angle of the inner (i) and outer (o)
belts respectively.

The radial optical depth of inner belt, $\tau_{\rm i}$, is related to the opening angle
by
\begin{equation}\label{eq:inner}
  f_{\rm i} = \tau_{\rm i} \sin( \theta_{\rm i} / 2)
\end{equation}
That is, the fraction of starlight captured by the inner belt, and hence the fractional
luminosity, is the radial optical depth multiplied by the fraction of the sky that it
covers as seen by the star. When it is optically thick, increasing the radial optical
depth does not increase the belt brightness (i.e. in equation \ref{eq:inner}, $\tau_{\rm
  i} \le 1$), so the opening angle of the inner belt must be at least 7$^\circ$. The
radial optical depth is correspondingly lower if the opening angle is larger, with a
lower a limit of 0.06 if the belt is a spherical shell (i.e. $\theta_{\rm i}=180^\circ$).

The outer belt is always shadowed by the inner belt to some degree. If the opening angle
of the outer belt is less than that of the inner belt, all light illuminating the outer
belt must pass through the inner belt, with the light attenuated by a factor $1-\tau_{\rm
  i}$ (with a minimum value of zero). If the inner belt is fatter than the outer belt,
then only a fraction of the light heating the outer belt is intercepted by the inner
belt, and the outer belt thickness is independent of the inner belt thickness (for fixed
$f_{\rm i}$). In these two regimes, the radial optical depth of the outer belt,
$\tau_{\rm o}$ (also $\le$1), is related to its opening angle and the inner belt
properties by
\begin{equation}\label{eq:outer}
  f_{\rm o} = \left\{ \begin{array}{ll}
      \tau_{\rm o} (1-\tau_{\rm i}) \sin( \theta_{\rm o}/2) & \theta_{\rm o} < \theta_{\rm i} \\
      \tau_{\rm o} \left( \sin( \theta_{\rm 0}/2) - f_{\rm i} \right) & \theta_{\rm o} > \theta_{\rm i}
    \end{array}
  \right.
\end{equation}

These relations are shown by the solid lines in Figure \ref{fig:opang}, which uses the
parameter space of $\theta_{\rm o}$ vs. $\theta_{\rm i}$. Dashed lines show constant
inner belt optical depth, where the minimum inner belt opening angle is set by $\tau_{\rm
  i} = 1$. The solid lines show equation (\ref{eq:outer}) for three different inner belt
optical depths. There are two different regimes for each line; if $\theta_{\rm o} <
\theta_{\rm i}$ the outer belt is fully shadowed by the inner and there is a tradeoff
between the opening angles of each, the thinner the inner belt the more it shadows the
outer belt due to increased optical depth, and the thicker the outer belt must be. This
regime is shown by the curved part of the solid lines. If $\theta_{\rm o} > \theta_{\rm
  i}$ the outer belt is thicker than the inner and changes in the inner belt thickness do
not change the outer belt brightness. That is, a thicker inner belt occults the outer
belt more, but is less optically thick for the same fractional luminosity. The $\tau_{\rm
  o}=1$ line tends towards, but never reaches $2\sin^{-1}f_{\rm o} \approx 5^\circ$ for
large $\theta_{\rm i}$, which would be the minimum opening angle in the absence of the
inner belt.

Therefore, if both belts are to be relatively thin the minimum opening angles allowed by
this simple analysis are about 7$^\circ$ for the inner component, and $\sim$12$^\circ$
for the outer belt (i.e. the left end of the lowest solid line). This minimum assumes
that both components are radially optically thick, and that no light is lost to
scattering, the latter being a reasonable approximation given that typical albedoes for
small bodies in the Solar System and in other debris discs are $\sim$10\%
\citep[e.g.][]{2005Natur.435.1067K,2008DPS....40.6108M,2009AJ....138..240F,2010AJ....140.1051K}. The
other two solid lines show that even for larger opening angles the belts must have
relatively large radial optical depths, purely due to the high fractional luminosity.

There are few debris discs where the opening angles can be measured, though a wealth of
information is available for the Solar System. The inclinations of Asteroid belt objects
vary widely, with most objects below 10$^\circ$ (an opening angle of 20$^\circ$), and
similar values for Kuiper belt objects. Similarly, the full-width at half-maximum
brightness opening angle of the $\beta$ Pictoris disc is $\sim$10$^\circ$ at about 130AU
\citep{2006AJ....131.3109G}, and the AU Microscopii disc is even thinner at
$\sim$5$^\circ$ \citep{2005AJ....129.1008K}. Therefore, the opening angles inferred for
the inner and outer HD~166191 disc components are comparable with observed
discs. However, this similarity requires that both components are radially optically
thick \emph{and} optically thin in the viewing direction, or the opening angles must be
larger. For a typically inclined viewing geometry ($\sim$60$^\circ$) such a configuration
is unlikely, so unless the disc components are much more radially extended than they are
vertically, and the system is viewed face-on, the opening angles are probably larger then
our estimated minimum. The star is not seen to be reddened, and given the large optical
depths seen in Figure \ref{fig:opang} the star is very probably seen directly and not
obscured by the disc. In summary, this comparison is suggestive that the two-belt debris
model requires relatively large optical depths for reasonable opening angles, but does
not strongly argue against such an interpretation.

Now considering the plausibility of the debris disc interpretation in terms of a
collisional model, the central issue is the lifetime of such extreme levels given the
$\sim$5~Myr age of HD~166191. As noted by \citet{2013arXiv1308.0405S}, the dust lifetime
due to collisions at about 1AU is only a few years so they must be replenished; a
late-stage planetary embryo impact or ongoing collisions in a massive Asteroid-belt
analogue are the possible sources. While either the inner or outer disc component could
be the result of a collision analogous to the Earth-Moon forming event, the extremely
bright period that follows such a collision is probably short lived. The length of this
period is uncertain due to the possibility of non-axisymmetry and an inability to remove
dust at high optical depth, but given the rapid evolution shown in Figure 14 of
\citet{2012MNRAS.tmp.3462J}, less than 100-1000 years seems a reasonable estimate. If the
outer belt is several times more distant the emission will decay more slowly, by roughly
a few orders of magnitude. Therefore, for a 5~Myr old star it may be unlikely to observe
short lived bright dust emission. However, this star was of course picked from a large
sample of stars due to its extremely bright excess and several giant collisions are
expected during terrestrial planet building
\citep[e.g.][]{1998Icar..136..304C,2004Icar..168....1R,2006AJ....131.1837K}

The alternative scenario suggested by \citet{2013arXiv1308.0405S} is dust resulting from
ongoing collisions in a massive Asteroid belt. The issue here is the same as above; such
extreme brightness requires a significant mass in large objects to generate a large
surface area of dust, but this mass (and hence space density) also requires that the
collisions between the large objects are very frequent. Indeed,
\citet{2007ApJ...658..569W} showed using a simple collision model that the maximum
fractional luminosity for young stars is $\sim$0.1\% (see their Figure 3). Alternatively,
using equation 21 of \citet{2007ApJ...658..569W}, to maintain a fractional luminosity
above 6\%, the maximum time the inner belt has been evolving for is of order 1,000 years,
with a few orders of magnitude uncertainty \citep[see also][]{2010MNRAS.401..867H}.

Another consideration for the massive Asteroid-belt analogue scenario is the disc
mass. \citet{2013arXiv1308.0405S} estimate the mass in dust present in the inner belt to
be roughly $10^{20}$ to $10^{23}$g. However, because the debris disc interpretation
explicitly implies that this mass is replenished by destruction of larger objects, the
dust mass is a negligible fraction of the total mass for typical debris disc size
distributions \citep[e.g.][]{1969JGR....74.2531D,2003Icar..164..334O}. Using equation 15
from \citet{2008ARA&A..46..339W} we can calculate the total mass present in a size
distribution from 1~$\mu$m sized grains up to 100km objects using a standard differential
size distribution slope of -3.5. Converting the 760K and 175K temperatures of the belts
\citep{2013arXiv1308.0405S} to radii of 0.3AU and 6AU, and using the fractional
luminosities of 0.06 and 0.04 yields total masses of 1.7 and 400$M_\oplus$ for the inner
and outer belts respectively. For reference, the minimum mass contained in the Solar
nebula was a few hundred Earth masses \citep[e.g.][]{1977Ap&SS..51..153W} so the inferred
mass for the outer belt is very large, if relatively large planetesimals are assumed. For
our assumed $dn(a) \propto a^{-3.5} da$ particle size distribution, the total mass scales
as the square root of the maximum size, so for much smaller 10m planetesimals, the mass
is also much smaller at $4M_\oplus$. However, the collisional lifetime calculated above
is also a function of planetesimal size ($\propto \sqrt{D}$) and assumed $D=2000$km, so
the outer belt lifetime for such small planetesimals would be of order 1000
years. Therefore, there is a trade-off between planetesimal size and disc lifetime,
meaning that either the disc is relatively massive, or its lifetime at the curent
brightness is very short.

In summary, we cannot rule out the debris disc interpretation of
\citet{2013arXiv1308.0405S}. We have shown i) that dust replenished by collisions either
requires near optically thick belts or relatively large disc opening angles and ii) that
the outer belt is relatively massive if it is the result of ongoing collisions or has a
very short lifetime. HD~166191 was selected as potentially interesting from a parent
population of $\approx$24,000 Hipparcos stars \citep{2013MNRAS.433.2334K} and is
therefore a rare object within this sample, meaning that arguments against the debris
disc interpretation that rely on HD~166191 being observed at a special time may not be
robust. As we have noted however, there is another possible interpretation; a
protoplanetary disc made up of both gas and dust, as is seen around many stars that have
similar ages to HD~166191.

\subsection{A transition disc?}\label{ss:rt}

\begin{figure}
  \begin{center}
    \hspace{-0.5cm} \includegraphics[width=0.5\textwidth]{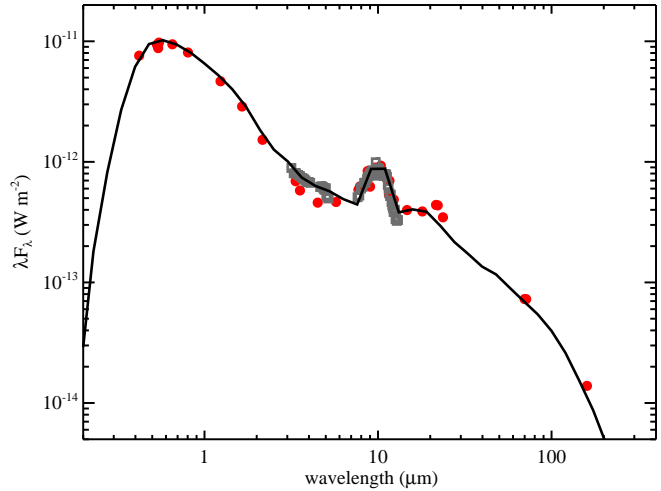}
    \caption{Protoplanetary transition disc model for HD~166191. Red dots show photometry
      and grey squares the BASS spectrum. The black solid line shows the best fitting
      model as described in the text.}\label{fig:rtmod}
  \end{center}
\end{figure}

To test whether HD~166191 can be modelled as a protoplanetary disc we modelled the
observed spectral energy distribution (SED) using the Monte Carlo Radiative transfer code
MCFOST \citep{2006A&A...459..797P,2009A&A...498..967P}. Here, we used a parametric disc
extending from an inner radius $R_{\rm in}$ to an outer radius $R_{\rm out}$. With this
approach, the model is of a dust disc that is implicitly well mixed with the gas that
makes up most of the total mass in the disc. The dust density distribution has a Gaussian
vertical profile, and the dust surface density profile and the scale height are defined
as power-law distributions with $\rho (r, z) = \rho_0 (r) \exp(-{z^2} / 2{h^2}(r)),
\Sigma (r) = \Sigma_0 (r / r_0)^{-p}$ and $ h(r) = h_0 (r / r_0)^{\beta}$, respectively,
where $r$ is the radial coordinate in the equatorial plane and $h_0$ is the scale height
at a fiducial radius $r_0$ = 100 AU. The dust grains are assumed to be homogeneous and
spherical amorphous silicates with a differential grain size distribution of the form
$dn(a) \propto a^{-3.5}da$, between a minimum grain size $a_{\rm min}$ and a maximum
grain size $a_{\rm max}$. The dust extinction and scattering opacities, scattering phase
functions, and Mueller matrices are calculated using Mie theory.

In order to fit the SED data, we first searched for a satisfactory solution by hand,
which was used as the starting point for an automated best fit search with a Genetic
Algorithm. In the absence of strong constraints from direct imaging (the \emph{Herschel}
observations require $R_{\rm out} \lesssim 300$ AU), we have set the external radius
$R_{\rm out}= 25$ AU. This value is not well constrained by the SED modelling, in
particular because of the lack of photometric data longer than 160 $\mu$m. The parameters
of the best fitting solution are presented in Table \ref{tab:parameters} and the
resulting SED is presented in Figure \ref{fig:rtmod}. The fit to the data is sufficiently
good to show that the HD~166191 spectrum can be interpreted as a protoplanetary disc.

A prominent feature of the SED is the 10 $\mu$m silicate emission, though we did not
thoroughly explore the mineralogical composition of the disc. However, to better match
the general profile of the 10 $\mu$m feature we added a fraction of crystalline silicates
\citep{2001ApJ...550L.213L} to the dust mixture. The crystalline silicates have a size
distribution ranging from 0.03 to 5 $\mu$m. The amorphous silicates \citep[of olivine
stoichiometry,][]{1995A&A...300..503D} have a size distribution ranging form 0.03 to 9
$\mu$m. This value for the maximum size of the amorphous silicates is poorly constrained
because of the lack of data longer than 160 $\mu$m. Both species are distributed
similarly in the disc. The Genetic Algorithm found the best model for a composition made
of 18\% of crystalline silicates, similar to the fraction of warm crystalline silicates
found by \citet{2010A&A...520A..39O} in a study of 58 T Tauri stars with IRS spectra.

The resulting model requires the scale height to be about 4 AU at $r=100$ AU (or
$h/r=0.04$). This is smaller than the ``standard'' protoplanetary disc models that
usually have $h/r \sim 0.1$, indicating that the dust has settled towards the midplane
\citep[e.g.][]{2011ApJS..193...11M,2011ApJ...742...39S}. The derived inner radius of
0.95~AU suggests that the disc has an inner hole, which along with the lack of detectable
accretion (\S \ref{ss:sso}) is consistent with a several Myr old object that has started
the transition from a full protoplanetary disc to a debris disc. The inner hole size is
smaller than most transition discs, but a wide range of sizes is seen
\citep[e.g.][]{2012ApJ...747..103E,2011ApJ...732...42A} so a $\sim$1~AU hole is not
particularly remarkable. The far-IR disc emission level is the main indicator of
settling, which can be significantly brighter in other transition discs (see fig
\ref{fig:speccomp} below), suggesting that the progression towards the debris disc phase
is more advanced for HD~166191 \citep[e.g.][]{2011ApJ...742...39S}.

\begin{table}
  \begin{center}
    \caption{Parameters of the best-fit model}
    \label{tab:parameters}
    \begin{tabular}{lcc}
      \hline\hline
      Parameter & explored & best\\
      & range & value  \\
      \hline
      \multicolumn{3}{c}{Stellar parameters} \\
      \hline
      $T_{\rm eff}$ (K) & fixed  & 6000 \\
      $R_{\rm star}$ ($R_\odot$) & fixed & 2.13 \\
      $A_V$ & fixed & 0.0 \\
      Distance (pc) & fixed &  119 \\
      \hline
      \multicolumn{3}{c}{Disc parameters} \\
      \hline
      M$_{\rm dust}$  ($M_\odot$) & 10$^{-7}$ -- 10$^{-4}$ & $5 \times 10^{-6}$ \\
      $R_{\rm out}$ (AU) & fixed & 25   \\
      $R_{\rm in}$ (AU)  & 0.1 -- 5.0 & 0.95 \\
      $\beta$ & 1.0 -- 1.3 & 1.12 \\ 
      p & -2.0 --  -0.1 & -0.8 \\
      h$_0$ (AU, \@r=100AU) & 1.0 -- 15.0 & 3.8 \\
      inclin. (deg.) & 0.0 -- 90.0 & $<60$ \\
      a$_{\rm min}$ ($\mu$m) & fixed  &  0.03  \\ 
      a$_{\rm max-amorph}$ ($\mu$m) & 0.5 -- 1000.0  &  9.0  \\ 
      Amorph. to Cryst. ratio & 1.0 -- 0.7 & 0.82 \\
      \hline
    \end{tabular}
  \end{center}
\end{table}

\section{Discussion}\label{s:disc}

Perhaps the biggest clue to the status of HD~166191 is the common space motion with the
well known Herbig Ae star HD~163296. As outlined in \S \ref{s:age}, this association
suggests that HD~166191 has a similar $\sim$4-5~Myr age to HD 163296, and these stars may
represent the first few members of a yet-to-be discovered association or
Scorpius-Centaurus subgroup that comprises many more lower-mass stars.

The implication for HD~166191 itself is that the star probably hosts a protoplanetary
disc rather than an exceptionally bright debris disc. The young age does not of course
require that the disc is primordial, but places HD~166191 in the $<$10~Myr age range
where Sun-like stars are seen to host such discs
\citep[e.g.][]{2007ApJ...667..308C,2009ApJS..181..197C,2011ARA&A..49...67W}. While we
cannot rule out the extreme debris disc interpretation, we have noted potential issues
(\S \ref{ss:2b}), and suggest that the disc is actually gas-rich. Our protoplanetary disc
interpretation is corroborated by the fact that the disc spectrum is consistent with a
fairly standard model for a protoplanetary disc (see \S \ref{ss:rt}). The spectrum shows
evidence for clearing in the innermost regions, as expected given typical evolutionary
sequences, and grain settling in the outer regions, both signatures of discs whose
evolution is well advanced beyond their initial state, as would be expected for a
$\sim$5~Myr old star.

\begin{figure}
  \begin{center}
    \hspace{-0.5cm} \includegraphics[width=0.5\textwidth]{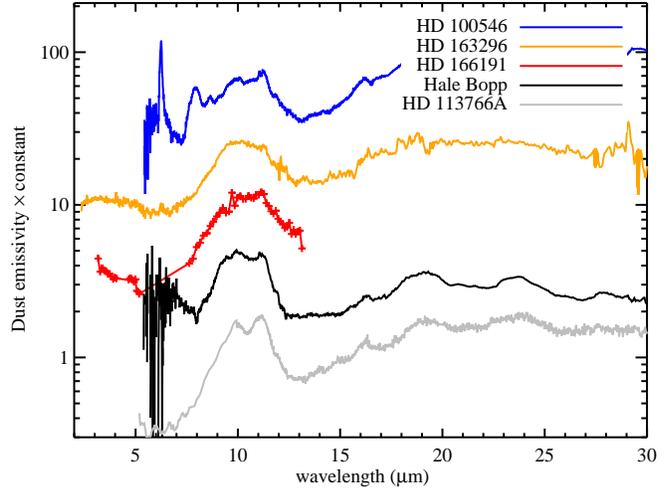}
    \caption{Comparison of the mid-IR emissivity of HD~166191 with other protoplanetary
      (HD~100546 and HD~163296), and debris (HD~113766A) discs, and comet Hale Bopp. The
      emissivity is calculated by dividing the spectrum by a blackbody at the approximate
      temperature of the continuum emission
      \citep[e.g.][]{2008ApJ...673.1106L}.}\label{fig:midircomp}
  \end{center}
\end{figure}

Further clues to the nature of the disc lie in the BASS spectrum, which is compared to
other discs and comet Hale Bopp's coma in Figure \ref{fig:midircomp}. Here we have
divided each spectrum by a blackbody at the appropriate temperature to produce
``emissivity'' spectra, which far better show spectral features common to objects
emitting at different temperatures.\footnote{Using the temperatures of 250 \& 135K
  \citep[HD~100546,][]{2007Icar..187...69L}, 300K (HD~163296), 700K (HD~166191), 200K
  \citep[Hale Bopp,][]{2007Icar..187...69L}, and 440K
  \citep[HD~113766A,][]{2008ApJ...673.1106L}.}  Of particular note here is the similarity
between the HD~166191 and HD~163296 peaks around 10~$\mu$m. Compared to HD~113766A,
another candidate for ongoing terrestrial planet formation, the HD~166191 silicate peak
is broader and is not double peaked.

Such differences in the shape of the silicate feature are diagnostic of the grain sizes
and composition, which can subsequently be related to their evolutionary state. One
approach is to derive compositions for objects of a range of age via modelling, and then
comparing the results in terms of an evolutionary sequence. For example,
\citet{2008ApJ...673.1106L} showed that their derived values for the olivine to pyroxene
ratio change with age. \citet{2012ApJ...747...93L} show the olivine to pyroxene ratio and
total olivine fraction derived from modelling for 19 objects in their Figure 7. Older
objects have greater olivine fractions, both in absolute terms and relative to pyroxene,
with HD~163296 appearing as one of the least evolved objects and HD~113766A appearing as
a moderately evolved object. Therefore, because the spectrum of HD~166191 appears very
similar in shape to HD~163296, and different to HD~113766A, the evolutionary state of the
HD~166191 dust appears more primordial. This conclusion would of course be strengthened
by detailed modeling of a 5-35 $\mu$m spectrum, as was done for the other systems in Fig
7, which allows for verification of the 8-13 $\mu$m silicate identifications using the
secondary 16 - 33 $\mu$m silicate features. We thus look forward to obtaining a detailed
spectrum of the HD166191 system in the 13 - 30 $\mu$m range, e.g. with JWST.

\begin{figure}
  \begin{center}
    \hspace{-0.5cm} \includegraphics[width=0.5\textwidth]{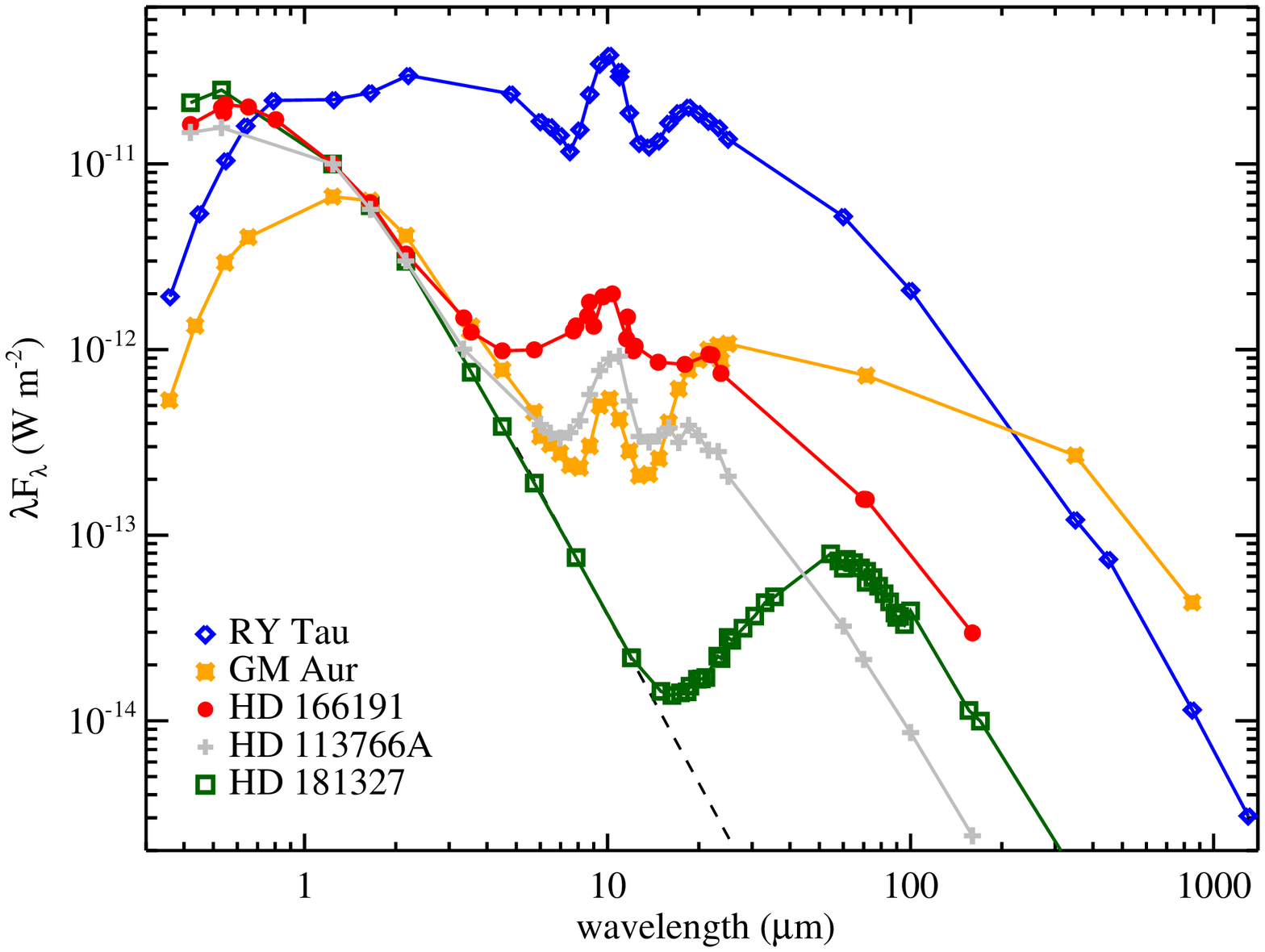}
    \caption{Comparison of HD~166191 with other protoplanetary (RY~Tau, GM~Aur) and
      debris (HD~113766A, HD~181327) discs. Data compiled from: RY Tau, GM Aur
      \citep{2007ApJS..169..328R}, HD~113766A
      \citep{2008ApJ...673.1106L,2013A&A...551A.134O}, and HD~181327
      \citep{2012A&A...539A..17L}, using \emph{Spitzer} InfraRed Spectrograph
      \citep[IRS,][]{2004ApJS..154...18H} spectra from the CASSIS database
      \citep{2011ApJS..196....8L}.}\label{fig:speccomp}
  \end{center}
\end{figure}

The overall disc spectrum should also be put in context with entire disc spectra at other
stages to better picture HD~166191 as part of a possible evolutionary sequence. Figure
\ref{fig:speccomp} compares the overall 0.5 - 160 $\mu$m HD~166191 SED with some examples
of other protoplanetary and debris discs around (approximately) Sun-like stars, all
scaled to a similar photospheric level at 2$\mu$m, with the exception of RY~Tau for which
the photosphere is not visible at this wavelength. We have not attempted to correct the
shorter wavelength photometry for reddening as we are here concerned with the longer
wavelengths. These represent something of an evolutionary sequence; (i) the youngest
discs, represented here by RY~Tau \citep{2008A&A...478..779S}, have a relatively flat
spectral slope across near/mid-IR wavelengths, but at the onset of disc dispersal (ii) an
inner hole develops and the spectrum has a flux deficit in the mid-IR and little change
in the far-IR and sub-millimetre flux \citep[e.g. GM~Aur,][]{2005ApJ...630L.185C}, the
final debris disc phase (iii) begins when the gas disc is dispersed, and generally the
disc has a spectrum that is typically very similar to a pure blackbody
\citep[e.g. HD~181327,][]{2012A&A...539A..17L}, but in some rare cases shows significant
levels of warm emission, which is commonly interpreted as a sign of terrestrial planet
formation \citep[e.g. HD~113766A,][]{2008ApJ...673.1106L,2013A&A...551A.134O}. Both
protoplanetary and debris discs show solid-state 10~$\mu$m silicate features, so their
presence or otherwise does not indicate the evolutionary state in this broad comparison
(though as discussed above their shape is important).

Within the context of this simple evolutionary sequence, HD~166191 shows characteristics
that might individually be attributed to either a transitional or a debris disc. The disc
shows evidence for an inner hole with little near-IR dust emission, though the hole is
probably not as large or well cleared as for GM Aur given the larger mid-IR excess. In an
inside-out evolution picture such as photoevaporation
\citep[e.g.][]{2001MNRAS.328..485C}, the HD~166191 disc might be considered less evolved
than GM Aur in the inner regions. However, as we note below it may be that the inner
emission is being enhanced by dust produced during terrestrial planet formation, which
could happen before and/or after an inner hole develops. A significant difference with GM
Aur is a lower level of far-IR and sub-millimetre excess, suggesting that the HD~166191
disc is significantly less flared as a result of a lower disc mass and/or dust grains
settling to the disc midplane.

Indeed, comparison of the HD~166191 far-IR emission with HD~181327, a young ($\sim$10~Myr)
bright debris disc, shows that the two have similar levels of excess, and that both are
lower than ``typical'' protoplanetary discs such as RY Tau and GM Aur. This comparable
level of far-IR flux suggests that in addition to being less flared, that the HD~166191
disc could be well advanced in the transition to a debris disc in the outer
regions. Similarly evolved discs have been observed around lower mass stars
\citep[e.g. in Corona Australis,][]{2013A&A...551A..34S}, suggesting that such a spectrum
is not particularly unusual. For example, in the $F_{24}/F_{100}$ vs. $F_{24}$ diagram of
\citet{2013A&A...551A..34S} (their Figure 11), HD~166191 would lie among the ``Depleted
discs'', indicating significant evolution and lower scale heights. How far advanced the
transition to the debris disc phase is will require a gas detection or stringent limits.

Finally, comparing HD~166191 with HD~113766A, a bright warm debris disc, the 0.5 - 160
$\mu$m spectrum continua look very similar; HD~113766A plausibly looks like a somewhat
more evolved and fainter version of HD~166191. Given that HD~166191 can be modelled as a
transition disc, such a comparison suggests that the interpretation of HD~113766A, which
has thus far been classed as a debris disc,
\citep[e.g.][]{2008ApJ...673.1106L,2012MNRAS.422.2560S,2013A&A...551A.134O}, deserves
further consideration. For example, it may be that the two-component structure inferred
from the combination of mid-IR interferometry \citep{2012MNRAS.422.2560S} and far-IR
photometry \citep{2013A&A...551A.134O} is actually indicative of an inner disc that is
significantly extended, as might be expected for a protoplanetary disc. However,
\citet{2006ApJS..166..351C} set stringent limits on gas levels in the inner regions of
the HD~113766A disc using the IRS spectrum, and the dust emission may therefore not be
primordial without gas to protect it from radiation forces. In addition, we have already
argued that HD~113766A appears more evolved based on the mid-IR silicate feature (Figure
\ref{fig:midircomp}). This possibility leads us to a third interpretation for the
HD~166191 disc (that also applies to HD~113766A) that we have not yet considered; a
combination of scenarios where the outer disc is still gaseous with significantly settled
dust, while the inner regions are in the late stages of, or have already made, the
transition to the debris phase. In this case the excess might be again thought of as
comprising two components; an inner warm component due at least in part to ongoing
terrestrial planet formation (but perhaps also the remnants of the protoplanetary disc),
and a well settled outer component with residual gas. Remnant gas may help maintain dust
levels produced by collisions, reducing the timescale problems that arise in the debris
disc scenario, where dust is removed easily by radiation forces.

Following such a phase, it may be that large cool discs such as HD~181327 emerge as the
inner regions collide and decay, and are then observed around older stars as ``typical''
(i.e. longer lived) debris discs at radii of tens to hundreds of AU. There are several
possible scenarios for this sequence however. For example, HD~166191 could already host a
debris disc that lies inside the protoplanetary disc and is currently undetectable, but
will emerge as the protoplanetary disc continues to disperse. Alternatively, it may be
that the birth of a debris disc requires stirring of planetesimals to sufficient
velocities through formation of large protoplanets, so the HD~166191 spectrum may in the
future drop to much fainter levels, but then become brighter again once the debris phase
begins \citep[e.g.][]{2004AJ....127..513K,2008ApJ...672..558C}

The single most important difference between the scenarios we have discussed, and the
fundamental difference between our transition disc interpretation and the debris disc
interpretation of \citet{2013arXiv1308.0405S}, is the presence or absence of gas.  While
the BASS spectrum is more similar to protoplanetary discs and argues that the silicate
emission is relatively primordial, this signature comes from dust. A disc that is still
in transition to the debris disc phase will have some remaining gas, while a true debris
disc will have negligible gas. There is currently no evidence for gas in the HD~166191
disc, either from gas accretion onto the star nor directly from the disc (e.g. from
near-IR CO emission in the SPeX spectrum). Indeed, one apparent reason
\citet{2013arXiv1308.0405S} reject the $\sim$5~Myr age suggested by the CMD in favour of
an older 10-100~Myr age is based on the lack of H$\alpha$ emission. However, many young
stars with protoplanetary discs are not seen to accrete gas, particularly transition
discs \citep[e.g.][]{2010ApJ...710..597S} and accretion is not necessarily expected
during this phase either \citep[e.g.][]{2006MNRAS.369..229A}. Detection, or a strongly
constraining non-detection, of gas around HD~166191 is therefore an obvious way in which
the disc may be further characterised.

\section{Summary}

HD~166191 has been known to host an infrared excess for several decades, but the origin
of this excess was unclear. New surveys by the AKARI and WISE telescopes have led to
renewed interest, and suggestions that the excess is due to a rare bright debris disc
\citep{2013A&A...550A..45F,2013arXiv1308.0405S} or an evolved protoplanetary disc
\citet{2013MNRAS.433.2334K}.

Though we cannot rule out the debris disc interpretation, we argue that the excess is
most simply interpreted as a relatively normal protoplanetary transition disc. This
conclusion is supported by our findings that i) HD~166191 is very likely associated and
co-eval with the previously isolated $\sim$4~Myr old Herbig Ae star HD~163296, ii) the
disc spectrum can be modelled by a protoplanetary transition disc, and iii) the mid-IR
silicate feature is a good match to other young objects, such as HD~163296. This finding
may represent the first hint of a new nearby young stellar association or subgroup of
Scorpius-Centaurus.

By comparing HD~166191 with other known protoplanetary and debris discs, we argue that
the HD~166191 disc is more evolved than most transition discs, and suggest some possible
scenarios for the late stages of the protoplanetary to debris disc transition.

\section{Acknowledgements}

We thank the referee for a careful reading of the manuscript and comments that improved
the article. This work was supported by the European Union through ERC grant number
279973 (GMK \& MCW). SJM is supported through a Gliese Fellowship at the University of
Heidelberg's Zentrum f\"{u}r Astronomie. FM acknowledges support from the Millennium
Science Initiative (Chilean Ministry of Economy), through grant ``Nucleus
P10-022-F''. FED acknowledges support by NASA under Grant No. NNX12AL26G issued through
the Planetary Astronomy Program.

%\bibliography{../ref,../extras} \bibliographystyle{apj}

\begin{thebibliography}{111}
\expandafter\ifx\csname natexlab\endcsname\relax\def\natexlab#1{#1}\fi

\bibitem[{{Adams} {et~al.}(2004){Adams}, {Hollenbach}, {Laughlin}, \&
  {Gorti}}]{2004ApJ...611..360A}
{Adams}, F.~C., {Hollenbach}, D., {Laughlin}, G., \& {Gorti}, U. 2004, \apj,
  611, 360

\bibitem[{{Alecian} {et~al.}(2013){Alecian}, {Wade}, {Catala}, {Grunhut},
  {Landstreet}, {Bagnulo}, {B{\"o}hm}, {Folsom}, {Marsden}, \&
  {Waite}}]{2013MNRAS.429.1001A}
{Alecian}, E., {Wade}, G.~A., {Catala}, C., {Grunhut}, J.~H., {Landstreet},
  J.~D., {Bagnulo}, S., {B{\"o}hm}, T., {Folsom}, C.~P., {Marsden}, S., \&
  {Waite}, I. 2013, \mnras, 429, 1001

\bibitem[{{Alexander} {et~al.}(2006){Alexander}, {Clarke}, \&
  {Pringle}}]{2006MNRAS.369..229A}
{Alexander}, R.~D., {Clarke}, C.~J., \& {Pringle}, J.~E. 2006, \mnras, 369, 229

\bibitem[{{Anderson} \& {Francis}(2012)}]{2012AstL...38..331A}
{Anderson}, E. \& {Francis}, C. 2012, Astronomy Letters, 38, 331

\bibitem[{{Andrews} {et~al.}(2011){Andrews}, {Wilner}, {Espaillat}, {Hughes},
  {Dullemond}, {McClure}, {Qi}, \& {Brown}}]{2011ApJ...732...42A}
{Andrews}, S.~M., {Wilner}, D.~J., {Espaillat}, C., {Hughes}, A.~M.,
  {Dullemond}, C.~P., {McClure}, M.~K., {Qi}, C., \& {Brown}, J.~M. 2011, \apj,
  732, 42

\bibitem[{{Bary} {et~al.}(2009){Bary}, {Leisenring}, \&
  {Skrutskie}}]{2009ApJ...706L.168B}
{Bary}, J.~S., {Leisenring}, J.~M., \& {Skrutskie}, M.~F. 2009, \apjl, 706,
  L168

\bibitem[{{Bayliss} {et~al.}(2013){Bayliss}, {Zhou}, {Penev}, {Bakos},
  {Hartman}, {Jord{\'a}n}, {Mancini}, {Mohler}, {Suc}, {Rabus}, {B{\'e}ky},
  {Csubry}, {Buchhave}, {Henning}, {Nikolov}, {Cs{\'a}k}, {Brahm}, {Espinoza},
  {Noyes}, {Schmidt}, {Conroy}, {Wright}, {Tinney}, {Addison}, {Sackett},
  {Sasselov}, {L{\'a}z{\'a}r}, {Papp}, \& {S{\'a}ri}}]{2013arXiv1306.0624B}
{Bayliss}, D., {Zhou}, G., {Penev}, K., {Bakos}, G., {Hartman}, J.,
  {Jord{\'a}n}, A., {Mancini}, L., {Mohler}, M., {Suc}, V., {Rabus}, M.,
  {B{\'e}ky}, B., {Csubry}, Z., {Buchhave}, L., {Henning}, T., {Nikolov}, N.,
  {Cs{\'a}k}, B., {Brahm}, R., {Espinoza}, N., {Noyes}, R., {Schmidt}, B.,
  {Conroy}, P., {Wright}, D., {Tinney}, C., {Addison}, B., {Sackett}, P.,
  {Sasselov}, D., {L{\'a}z{\'a}r}, J., {Papp}, I., \& {S{\'a}ri}, P. 2013,
  ArXiv e-prints

\bibitem[{{Bergin} {et~al.}(2004){Bergin}, {Calvet}, {Sitko}, {Abgrall},
  {D'Alessio}, {Herczeg}, {Roueff}, {Qi}, {Lynch}, {Russell}, {Brafford}, \&
  {Perry}}]{2004ApJ...614L.133B}
{Bergin}, E., {Calvet}, N., {Sitko}, M.~L., {Abgrall}, H., {D'Alessio}, P.,
  {Herczeg}, G.~J., {Roueff}, E., {Qi}, C., {Lynch}, D.~K., {Russell}, R.~W.,
  {Brafford}, S.~M., \& {Perry}, R.~B. 2004, \apjl, 614, L133

\bibitem[{{Bessell}(1999)}]{1999PASP..111.1426B}
{Bessell}, M.~S. 1999, \pasp, 111, 1426

\bibitem[{{Bessell}(2000)}]{2000PASP..112..961B}
---. 2000, \pasp, 112, 961

\bibitem[{{Bressan} {et~al.}(2012){Bressan}, {Marigo}, {Girardi}, {Salasnich},
  {Dal Cero}, {Rubele}, \& {Nanni}}]{2012MNRAS.427..127B}
{Bressan}, A., {Marigo}, P., {Girardi}, L., {Salasnich}, B., {Dal Cero}, C.,
  {Rubele}, S., \& {Nanni}, A. 2012, \mnras, 427, 127

\bibitem[{{Brott} \& {Hauschildt}(2005)}]{2005ESASP.576..565B}
{Brott}, I. \& {Hauschildt}, P.~H. 2005, in ESA Special Publication, Vol. 576,
  The Three-Dimensional Universe with Gaia, ed. C.~{Turon}, K.~S. {O'Flaherty},
  \& M.~A.~C. {Perryman}, 565

\bibitem[{{Buscombe} \& {Kennedy}(1965)}]{1965MNRAS.130..281B}
{Buscombe}, W. \& {Kennedy}, P.~M. 1965, \mnras, 130, 281

\bibitem[{{Calvet} {et~al.}(2002){Calvet}, {D'Alessio}, {Hartmann}, {Wilner},
  {Walsh}, \& {Sitko}}]{2002ApJ...568.1008C}
{Calvet}, N., {D'Alessio}, P., {Hartmann}, L., {Wilner}, D., {Walsh}, A., \&
  {Sitko}, M. 2002, \apj, 568, 1008

\bibitem[{{Calvet} {et~al.}(2005){Calvet}, {D'Alessio}, {Watson},
  {Franco-Hern{\'a}ndez}, {Furlan}, {Green}, {Sutter}, {Forrest}, {Hartmann},
  {Uchida}, {Keller}, {Sargent}, {Najita}, {Herter}, {Barry}, \&
  {Hall}}]{2005ApJ...630L.185C}
{Calvet}, N., {D'Alessio}, P., {Watson}, D.~M., {Franco-Hern{\'a}ndez}, R.,
  {Furlan}, E., {Green}, J., {Sutter}, P.~M., {Forrest}, W.~J., {Hartmann}, L.,
  {Uchida}, K.~I., {Keller}, L.~D., {Sargent}, B., {Najita}, J., {Herter},
  T.~L., {Barry}, D.~J., \& {Hall}, P. 2005, \apjl, 630, L185

\bibitem[{{Carpenter} {et~al.}(2009){Carpenter}, {Bouwman}, {Mamajek}, {Meyer},
  {Hillenbrand}, {Backman}, {Henning}, {Hines}, {Hollenbach}, {Kim},
  {Moro-Martin}, {Pascucci}, {Silverstone}, {Stauffer}, \&
  {Wolf}}]{2009ApJS..181..197C}
{Carpenter}, J.~M., {Bouwman}, J., {Mamajek}, E.~E., {Meyer}, M.~R.,
  {Hillenbrand}, L.~A., {Backman}, D.~E., {Henning}, T., {Hines}, D.~C.,
  {Hollenbach}, D., {Kim}, J.~S., {Moro-Martin}, A., {Pascucci}, I.,
  {Silverstone}, M.~D., {Stauffer}, J.~R., \& {Wolf}, S. 2009, \apjs, 181, 197

\bibitem[{{Chambers} \& {Wetherill}(1998)}]{1998Icar..136..304C}
{Chambers}, J.~E. \& {Wetherill}, G.~W. 1998, \icarus, 136, 304

\bibitem[{{Chen} {et~al.}(2011){Chen}, {Mamajek}, {Bitner}, {Pecaut}, {Su}, \&
  {Weinberger}}]{2011ApJ...738..122C}
{Chen}, C.~H., {Mamajek}, E.~E., {Bitner}, M.~A., {Pecaut}, M., {Su}, K.~Y.~L.,
  \& {Weinberger}, A.~J. 2011, \apj, 738, 122

\bibitem[{{Chen} {et~al.}(2012){Chen}, {Pecaut}, {Mamajek}, {Su}, \&
  {Bitner}}]{2012ApJ...756..133C}
{Chen}, C.~H., {Pecaut}, M., {Mamajek}, E.~E., {Su}, K.~Y.~L., \& {Bitner}, M.
  2012, \apj, 756, 133

\bibitem[{{Chen} {et~al.}(2006){Chen}, {Sargent}, {Bohac}, {Kim},
  {Leibensperger}, {Jura}, {Najita}, {Forrest}, {Watson}, {Sloan}, \&
  {Keller}}]{2006ApJS..166..351C}
{Chen}, C.~H., {Sargent}, B.~A., {Bohac}, C., {Kim}, K.~H., {Leibensperger},
  E., {Jura}, M., {Najita}, J., {Forrest}, W.~J., {Watson}, D.~M., {Sloan},
  G.~C., \& {Keller}, L.~D. 2006, \apjs, 166, 351

\bibitem[{{Cieza} {et~al.}(2007)}]{2007ApJ...667..308C}
{Cieza}, L. {et~al.} 2007, \apj, 667, 308

\bibitem[{{Clarke} {et~al.}(2005){Clarke}, {Oudmaijer}, \&
  {Lumsden}}]{2005MNRAS.363.1111C}
{Clarke}, A.~J., {Oudmaijer}, R.~D., \& {Lumsden}, S.~L. 2005, \mnras, 363,
  1111

\bibitem[{{Clarke} {et~al.}(2001){Clarke}, {Gendrin}, \&
  {Sotomayor}}]{2001MNRAS.328..485C}
{Clarke}, C.~J., {Gendrin}, A., \& {Sotomayor}, M. 2001, \mnras, 328, 485

\bibitem[{{Currie} {et~al.}(2008){Currie}, {Kenyon}, {Balog}, {Rieke}, {Bragg},
  \& {Bromley}}]{2008ApJ...672..558C}
{Currie}, T., {Kenyon}, S.~J., {Balog}, Z., {Rieke}, G., {Bragg}, A., \&
  {Bromley}, B. 2008, \apj, 672, 558

\bibitem[{{Dohnanyi}(1969)}]{1969JGR....74.2531D}
{Dohnanyi}, J.~S. 1969, \jgr, 74, 2531

\bibitem[{{Dopita} {et~al.}(2007){Dopita}, {Hart}, {McGregor}, {Oates},
  {Bloxham}, \& {Jones}}]{2007Ap&SS.310..255D}
{Dopita}, M., {Hart}, J., {McGregor}, P., {Oates}, P., {Bloxham}, G., \&
  {Jones}, D. 2007, \apss, 310, 255

\bibitem[{{Dorschner} {et~al.}(1995){Dorschner}, {Begemann}, {Henning},
  {Jaeger}, \& {Mutschke}}]{1995A&A...300..503D}
{Dorschner}, J., {Begemann}, B., {Henning}, T., {Jaeger}, C., \& {Mutschke}, H.
  1995, \aap, 300, 503

\bibitem[{{Espaillat} {et~al.}(2011){Espaillat}, {Furlan}, {D'Alessio},
  {Sargent}, {Nagel}, {Calvet}, {Watson}, \& {Muzerolle}}]{2011ApJ...728...49E}
{Espaillat}, C., {Furlan}, E., {D'Alessio}, P., {Sargent}, B., {Nagel}, E.,
  {Calvet}, N., {Watson}, D.~M., \& {Muzerolle}, J. 2011, \apj, 728, 49

\bibitem[{{Espaillat} {et~al.}(2012){Espaillat}, {Ingleby}, {Hern{\'a}ndez},
  {Furlan}, {D'Alessio}, {Calvet}, {Andrews}, {Muzerolle}, {Qi}, \&
  {Wilner}}]{2012ApJ...747..103E}
{Espaillat}, C., {Ingleby}, L., {Hern{\'a}ndez}, J., {Furlan}, E., {D'Alessio},
  P., {Calvet}, N., {Andrews}, S., {Muzerolle}, J., {Qi}, C., \& {Wilner}, D.
  2012, \apj, 747, 103

\bibitem[{{Feigelson} \& {Montmerle}(1999)}]{1999ARA&A..37..363F}
{Feigelson}, E.~D. \& {Montmerle}, T. 1999, \araa, 37, 363

\bibitem[{{Fern{\'a}ndez} {et~al.}(2009){Fern{\'a}ndez}, {Jewitt}, \&
  {Ziffer}}]{2009AJ....138..240F}
{Fern{\'a}ndez}, Y.~R., {Jewitt}, D., \& {Ziffer}, J.~E. 2009, \aj, 138, 240

\bibitem[{{Fujiwara} {et~al.}(2013)}]{2013A&A...550A..45F}
{Fujiwara}, H. {et~al.} 2013, \aap, 550, A45

\bibitem[{{Furlan} {et~al.}(2007){Furlan}, {Sargent}, {Calvet}, {Forrest},
  {D'Alessio}, {Hartmann}, {Watson}, {Green}, {Najita}, \&
  {Chen}}]{2007ApJ...664.1176F}
{Furlan}, E., {Sargent}, B., {Calvet}, N., {Forrest}, W.~J., {D'Alessio}, P.,
  {Hartmann}, L., {Watson}, D.~M., {Green}, J.~D., {Najita}, J., \& {Chen},
  C.~H. 2007, \apj, 664, 1176

\bibitem[{{Golimowski} {et~al.}(2006){Golimowski}, {Ardila}, {Krist},
  {Clampin}, {Ford}, {Illingworth}, {Bartko}, {Ben{\'{\i}}tez}, {Blakeslee},
  {Bouwens}, {Bradley}, {Broadhurst}, {Brown}, {Burrows}, {Cheng}, {Cross},
  {Demarco}, {Feldman}, {Franx}, {Goto}, {Gronwall}, {Hartig}, {Holden},
  {Homeier}, {Infante}, {Jee}, {Kimble}, {Lesser}, {Martel}, {Mei},
  {Menanteau}, {Meurer}, {Miley}, {Motta}, {Postman}, {Rosati}, {Sirianni},
  {Sparks}, {Tran}, {Tsvetanov}, {White}, {Zheng}, \&
  {Zirm}}]{2006AJ....131.3109G}
{Golimowski}, D.~A., {Ardila}, D.~R., {Krist}, J.~E., {Clampin}, M., {Ford},
  H.~C., {Illingworth}, G.~D., {Bartko}, F., {Ben{\'{\i}}tez}, N., {Blakeslee},
  J.~P., {Bouwens}, R.~J., {Bradley}, L.~D., {Broadhurst}, T.~J., {Brown},
  R.~A., {Burrows}, C.~J., {Cheng}, E.~S., {Cross}, N.~J.~G., {Demarco}, R.,
  {Feldman}, P.~D., {Franx}, M., {Goto}, T., {Gronwall}, C., {Hartig}, G.~F.,
  {Holden}, B.~P., {Homeier}, N.~L., {Infante}, L., {Jee}, M.~J., {Kimble},
  R.~A., {Lesser}, M.~P., {Martel}, A.~R., {Mei}, S., {Menanteau}, F.,
  {Meurer}, G.~R., {Miley}, G.~K., {Motta}, V., {Postman}, M., {Rosati}, P.,
  {Sirianni}, M., {Sparks}, W.~B., {Tran}, H.~D., {Tsvetanov}, Z.~I., {White},
  R.~L., {Zheng}, W., \& {Zirm}, A.~W. 2006, \aj, 131, 3109

\bibitem[{{Gontcharov}(2006)}]{2006AstL...32..759G}
{Gontcharov}, G.~A. 2006, Astronomy Letters, 32, 759

\bibitem[{{Hackwell} {et~al.}(1990){Hackwell}, {Warren}, {Chatelain}, {Dotan},
  \& {Li}}]{1990SPIE.1235..171H}
{Hackwell}, J.~A., {Warren}, D.~W., {Chatelain}, M.~A., {Dotan}, Y., \& {Li},
  P.~H. 1990, in Society of Photo-Optical Instrumentation Engineers (SPIE)
  Conference Series, Vol. 1235, Society of Photo-Optical Instrumentation
  Engineers (SPIE) Conference Series, ed. D.~L. {Crawford}, 171--180

\bibitem[{{Haisch} {et~al.}(2001){Haisch}, {Lada}, \&
  {Lada}}]{2001ApJ...553L.153H}
{Haisch}, Jr., K.~E., {Lada}, E.~A., \& {Lada}, C.~J. 2001, \apjl, 553, L153

\bibitem[{{Hamuy} {et~al.}(1992){Hamuy}, {Walker}, {Suntzeff}, {Gigoux},
  {Heathcote}, \& {Phillips}}]{1992PASP..104..533H}
{Hamuy}, M., {Walker}, A.~R., {Suntzeff}, N.~B., {Gigoux}, P., {Heathcote},
  S.~R., \& {Phillips}, M.~M. 1992, \pasp, 104, 533

\bibitem[{{Heng} \& {Tremaine}(2010)}]{2010MNRAS.401..867H}
{Heng}, K. \& {Tremaine}, S. 2010, \mnras, 401, 867

\bibitem[{{Houck} {et~al.}(2004)}]{2004ApJS..154...18H}
{Houck}, J.~R. {et~al.} 2004, \apjs, 154, 18

\bibitem[{{Jackson} \& {Wyatt}(2012)}]{2012MNRAS.tmp.3462J}
{Jackson}, A.~P. \& {Wyatt}, M.~C. 2012, \mnras, 3462

\bibitem[{{Kalas} {et~al.}(2005){Kalas}, {Graham}, \&
  {Clampin}}]{2005Natur.435.1067K}
{Kalas}, P., {Graham}, J.~R., \& {Clampin}, M. 2005, \nat, 435, 1067

\bibitem[{{Kennedy} \& {Wyatt}(2013)}]{2013MNRAS.433.2334K}
{Kennedy}, G.~M. \& {Wyatt}, M.~C. 2013, \mnras, 433, 2334

\bibitem[{{Kenyon} \& {Bromley}(2004)}]{2004AJ....127..513K}
{Kenyon}, S.~J. \& {Bromley}, B.~C. 2004, \aj, 127, 513

\bibitem[{{Kenyon} \& {Bromley}(2006)}]{2006AJ....131.1837K}
---. 2006, \aj, 131, 1837

\bibitem[{{Krist} {et~al.}(2005){Krist}, {Ardila}, {Golimowski}, {Clampin},
  {Ford}, {Illingworth}, {Hartig}, {Bartko}, {Ben{\'{\i}}tez}, {Blakeslee},
  {Bouwens}, {Bradley}, {Broadhurst}, {Brown}, {Burrows}, {Cheng}, {Cross},
  {Demarco}, {Feldman}, {Franx}, {Goto}, {Gronwall}, {Holden}, {Homeier},
  {Infante}, {Kimble}, {Lesser}, {Martel}, {Mei}, {Menanteau}, {Meurer},
  {Miley}, {Motta}, {Postman}, {Rosati}, {Sirianni}, {Sparks}, {Tran},
  {Tsvetanov}, {White}, \& {Zheng}}]{2005AJ....129.1008K}
{Krist}, J.~E., {Ardila}, D.~R., {Golimowski}, D.~A., {Clampin}, M., {Ford},
  H.~C., {Illingworth}, G.~D., {Hartig}, G.~F., {Bartko}, F., {Ben{\'{\i}}tez},
  N., {Blakeslee}, J.~P., {Bouwens}, R.~J., {Bradley}, L.~D., {Broadhurst},
  T.~J., {Brown}, R.~A., {Burrows}, C.~J., {Cheng}, E.~S., {Cross}, N.~J.~G.,
  {Demarco}, R., {Feldman}, P.~D., {Franx}, M., {Goto}, T., {Gronwall}, C.,
  {Holden}, B., {Homeier}, N., {Infante}, L., {Kimble}, R.~A., {Lesser}, M.~P.,
  {Martel}, A.~R., {Mei}, S., {Menanteau}, F., {Meurer}, G.~R., {Miley}, G.~K.,
  {Motta}, V., {Postman}, M., {Rosati}, P., {Sirianni}, M., {Sparks}, W.~B.,
  {Tran}, H.~D., {Tsvetanov}, Z.~I., {White}, R.~L., \& {Zheng}, W. 2005, \aj,
  129, 1008

\bibitem[{{Krist} {et~al.}(2010){Krist}, {Stapelfeldt}, {Bryden}, {Rieke},
  {Su}, {Chen}, {Beichman}, {Hines}, {Rebull}, {Tanner}, {Trilling}, {Clampin},
  \& {G{\'a}sp{\'a}r}}]{2010AJ....140.1051K}
{Krist}, J.~E., {Stapelfeldt}, K.~R., {Bryden}, G., {Rieke}, G.~H., {Su},
  K.~Y.~L., {Chen}, C.~C., {Beichman}, C.~A., {Hines}, D.~C., {Rebull}, L.~M.,
  {Tanner}, A., {Trilling}, D.~E., {Clampin}, M., \& {G{\'a}sp{\'a}r}, A. 2010,
  \aj, 140, 1051

\bibitem[{{Lebouteiller} {et~al.}(2011){Lebouteiller}, {Barry}, {Spoon},
  {Bernard-Salas}, {Sloan}, {Houck}, \& {Weedman}}]{2011ApJS..196....8L}
{Lebouteiller}, V., {Barry}, D.~J., {Spoon}, H.~W.~W., {Bernard-Salas}, J.,
  {Sloan}, G.~C., {Houck}, J.~R., \& {Weedman}, D.~W. 2011, \apjs, 196, 8

\bibitem[{{Lebreton} {et~al.}(2012){Lebreton}, {Augereau}, {Thi}, {Roberge},
  {Donaldson}, {Schneider}, {Maddison}, {M{\'e}nard}, {Riviere-Marichalar},
  {Mathews}, {Kamp}, {Pinte}, {Dent}, {Barrado}, {Duch{\^e}ne}, {Gonzalez},
  {Grady}, {Meeus}, {Pantin}, {Williams}, \& {Woitke}}]{2012A&A...539A..17L}
{Lebreton}, J., {Augereau}, J.-C., {Thi}, W.-F., {Roberge}, A., {Donaldson},
  J., {Schneider}, G., {Maddison}, S.~T., {M{\'e}nard}, F.,
  {Riviere-Marichalar}, P., {Mathews}, G.~S., {Kamp}, I., {Pinte}, C., {Dent},
  W.~R.~F., {Barrado}, D., {Duch{\^e}ne}, G., {Gonzalez}, J.-F., {Grady},
  C.~A., {Meeus}, G., {Pantin}, E., {Williams}, J.~P., \& {Woitke}, P. 2012,
  \aap, 539, A17

\bibitem[{{Li} \& {Draine}(2001)}]{2001ApJ...550L.213L}
{Li}, A. \& {Draine}, B.~T. 2001, \apjl, 550, L213

\bibitem[{{Lisse} {et~al.}(2008){Lisse}, {Chen}, {Wyatt}, \&
  {Morlok}}]{2008ApJ...673.1106L}
{Lisse}, C.~M., {Chen}, C.~H., {Wyatt}, M.~C., \& {Morlok}, A. 2008, \apj, 673,
  1106

\bibitem[{{Lisse} {et~al.}(2007){Lisse}, {Kraemer}, {Nuth}, {Li}, \&
  {Joswiak}}]{2007Icar..187...69L}
{Lisse}, C.~M., {Kraemer}, K.~E., {Nuth}, J.~A., {Li}, A., \& {Joswiak}, D.
  2007, \icarus, 187, 69

\bibitem[{{Lisse} {et~al.}(2012){Lisse}, {Wyatt}, {Chen}, {Morlok}, {Watson},
  {Manoj}, {Sheehan}, {Currie}, {Thebault}, \& {Sitko}}]{2012ApJ...747...93L}
{Lisse}, C.~M., {Wyatt}, M.~C., {Chen}, C.~H., {Morlok}, A., {Watson}, D.~M.,
  {Manoj}, P., {Sheehan}, P., {Currie}, T.~M., {Thebault}, P., \& {Sitko},
  M.~L. 2012, \apj, 747, 93

\bibitem[{{Mamajek} {et~al.}(2000){Mamajek}, {Lawson}, \&
  {Feigelson}}]{2000ApJ...544..356M}
{Mamajek}, E.~E., {Lawson}, W.~A., \& {Feigelson}, E.~D. 2000, \apj, 544, 356

\bibitem[{{Mamajek} {et~al.}(2004){Mamajek}, {Meyer}, {Hinz}, {Hoffmann},
  {Cohen}, \& {Hora}}]{2004ApJ...612..496M}
{Mamajek}, E.~E., {Meyer}, M.~R., {Hinz}, P.~M., {Hoffmann}, W.~F., {Cohen},
  M., \& {Hora}, J.~L. 2004, \apj, 612, 496

\bibitem[{{Manoj} {et~al.}(2011)}]{2011ApJS..193...11M}
{Manoj}, P. {et~al.} 2011, \apjs, 193, 11

\bibitem[{{Melis} {et~al.}(2010){Melis}, {Zuckerman}, {Rhee}, \&
  {Song}}]{2010ApJ...717L..57M}
{Melis}, C., {Zuckerman}, B., {Rhee}, J.~H., \& {Song}, I. 2010, \apjl, 717,
  L57

\bibitem[{{Melis} {et~al.}(2012){Melis}, {Zuckerman}, {Rhee}, {Song}, {Murphy},
  \& {Bessell}}]{2012Natur.487...74M}
{Melis}, C., {Zuckerman}, B., {Rhee}, J.~H., {Song}, I., {Murphy}, S.~J., \&
  {Bessell}, M.~S. 2012, \nat, 487, 74

\bibitem[{{Meng} {et~al.}(2012){Meng}, {Rieke}, {Su}, {Ivanov}, {Vanzi}, \&
  {Rujopakarn}}]{2012ApJ...751L..17M}
{Meng}, H.~Y.~A., {Rieke}, G.~H., {Su}, K.~Y.~L., {Ivanov}, V.~D., {Vanzi}, L.,
  \& {Rujopakarn}, W. 2012, \apjl, 751, L17

\bibitem[{{Merrill}(1930)}]{1930ApJ....72...98M}
{Merrill}, P.~W. 1930, \apj, 72, 98

\bibitem[{{Moffett} \& {Barnes}(1984)}]{1984ApJS...55..389M}
{Moffett}, T.~J. \& {Barnes}, III, T.~G. 1984, \apjs, 55, 389

\bibitem[{{Mueller} {et~al.}(2008){Mueller}, {Grav}, {Trilling}, {Stansberry},
  \& {Sykes}}]{2008DPS....40.6108M}
{Mueller}, M., {Grav}, T., {Trilling}, D., {Stansberry}, J., \& {Sykes}, M.
  2008, in Bulletin of the American Astronomical Society, Vol.~40, AAS/Division
  for Planetary Sciences Meeting Abstracts \#40, 511

\bibitem[{{Murphy} {et~al.}(2013){Murphy}, {Lawson}, \&
  {Bessell}}]{2013MNRAS.435.1325M}
{Murphy}, S.~J., {Lawson}, W.~A., \& {Bessell}, M.~S. 2013, \mnras, 435, 1325

\bibitem[{{Muzerolle} {et~al.}(2010){Muzerolle}, {Allen}, {Megeath},
  {Hern{\'a}ndez}, \& {Gutermuth}}]{2010ApJ...708.1107M}
{Muzerolle}, J., {Allen}, L.~E., {Megeath}, S.~T., {Hern{\'a}ndez}, J., \&
  {Gutermuth}, R.~A. 2010, \apj, 708, 1107

\bibitem[{{Muzerolle} {et~al.}(2001){Muzerolle}, {Calvet}, \&
  {Hartmann}}]{2001ApJ...550..944M}
{Muzerolle}, J., {Calvet}, N., \& {Hartmann}, L. 2001, \apj, 550, 944

\bibitem[{{Muzerolle} {et~al.}(2009){Muzerolle}, {Flaherty}, {Balog}, {Furlan},
  {Smith}, {Allen}, {Calvet}, {D'Alessio}, {Megeath}, {Muench}, {Rieke}, \&
  {Sherry}}]{2009ApJ...704L..15M}
{Muzerolle}, J., {Flaherty}, K., {Balog}, Z., {Furlan}, E., {Smith}, P.~S.,
  {Allen}, L., {Calvet}, N., {D'Alessio}, P., {Megeath}, S.~T., {Muench}, A.,
  {Rieke}, G.~H., \& {Sherry}, W.~H. 2009, \apjl, 704, L15

\bibitem[{{Muzerolle} {et~al.}(2003){Muzerolle}, {Hillenbrand}, {Calvet},
  {Brice{\~n}o}, \& {Hartmann}}]{2003ApJ...592..266M}
{Muzerolle}, J., {Hillenbrand}, L., {Calvet}, N., {Brice{\~n}o}, C., \&
  {Hartmann}, L. 2003, \apj, 592, 266

\bibitem[{{O'Brien} \& {Greenberg}(2003)}]{2003Icar..164..334O}
{O'Brien}, D.~P. \& {Greenberg}, R. 2003, \icarus, 164, 334

\bibitem[{{Olofsson} {et~al.}(2010){Olofsson}, {Augereau}, {van Dishoeck},
  {Mer{\'{\i}}n}, {Grosso}, {M{\'e}nard}, {Blake}, \&
  {Monin}}]{2010A&A...520A..39O}
{Olofsson}, J., {Augereau}, J.-C., {van Dishoeck}, E.~F., {Mer{\'{\i}}n}, B.,
  {Grosso}, N., {M{\'e}nard}, F., {Blake}, G.~A., \& {Monin}, J.-L. 2010, \aap,
  520, A39

\bibitem[{{Olofsson} {et~al.}(2013){Olofsson}, {Henning}, {Nielbock},
  {Augereau}, {Juh{\`a}sz}, {Oliveira}, {Absil}, \&
  {Tamanai}}]{2013A&A...551A.134O}
{Olofsson}, J., {Henning}, T., {Nielbock}, M., {Augereau}, J.-C., {Juh{\`a}sz},
  A., {Oliveira}, I., {Absil}, O., \& {Tamanai}, A. 2013, \aap, 551, A134

\bibitem[{{Oudmaijer} {et~al.}(1992){Oudmaijer}, {van der Veen}, {Waters},
  {Trams}, {Waelkens}, \& {Engelsman}}]{1992A&AS...96..625O}
{Oudmaijer}, R.~D., {van der Veen}, W.~E.~C.~J., {Waters}, L.~B.~F.~M.,
  {Trams}, N.~R., {Waelkens}, C., \& {Engelsman}, E. 1992, \aaps, 96, 625

\bibitem[{{Owen} {et~al.}(2011){Owen}, {Ercolano}, \&
  {Clarke}}]{2011MNRAS.412...13O}
{Owen}, J.~E., {Ercolano}, B., \& {Clarke}, C.~J. 2011, \mnras, 412, 13

\bibitem[{{Penev} {et~al.}(2013){Penev}, {Bakos}, {Bayliss}, {Jord{\'a}n},
  {Mohler}, {Zhou}, {Suc}, {Rabus}, {Hartman}, {Mancini}, {B{\'e}ky}, {Csubry},
  {Buchhave}, {Henning}, {Nikolov}, {Cs{\'a}k}, {Brahm}, {Espinoza}, {Conroy},
  {Noyes}, {Sasselov}, {Schmidt}, {Wright}, {Tinney}, {Addison},
  {L{\'a}z{\'a}r}, {Papp}, \& {S{\'a}ri}}]{2013AJ....145....5P}
{Penev}, K., {Bakos}, G.~{\'A}., {Bayliss}, D., {Jord{\'a}n}, A., {Mohler}, M.,
  {Zhou}, G., {Suc}, V., {Rabus}, M., {Hartman}, J.~D., {Mancini}, L.,
  {B{\'e}ky}, B., {Csubry}, Z., {Buchhave}, L., {Henning}, T., {Nikolov}, N.,
  {Cs{\'a}k}, B., {Brahm}, R., {Espinoza}, N., {Conroy}, P., {Noyes}, R.~W.,
  {Sasselov}, D.~D., {Schmidt}, B., {Wright}, D.~J., {Tinney}, C.~G.,
  {Addison}, B.~C., {L{\'a}z{\'a}r}, J., {Papp}, I., \& {S{\'a}ri}, P. 2013,
  \aj, 145, 5

\bibitem[{{Pilbratt} {et~al.}(2010)}]{2010A&A...518L...1P}
{Pilbratt}, G.~L. {et~al.} 2010, \aap, 518, L1

\bibitem[{{Pinte} {et~al.}(2009){Pinte}, {Harries}, {Min}, {Watson},
  {Dullemond}, {Woitke}, {M{\'e}nard}, \&
  {Dur{\'a}n-Rojas}}]{2009A&A...498..967P}
{Pinte}, C., {Harries}, T.~J., {Min}, M., {Watson}, A.~M., {Dullemond}, C.~P.,
  {Woitke}, P., {M{\'e}nard}, F., \& {Dur{\'a}n-Rojas}, M.~C. 2009, \aap, 498,
  967

\bibitem[{{Pinte} {et~al.}(2006){Pinte}, {M{\'e}nard}, {Duch{\^e}ne}, \&
  {Bastien}}]{2006A&A...459..797P}
{Pinte}, C., {M{\'e}nard}, F., {Duch{\^e}ne}, G., \& {Bastien}, P. 2006, \aap,
  459, 797

\bibitem[{{Poglitsch} {et~al.}(2010)}]{2010A&A...518L...2P}
{Poglitsch}, A. {et~al.} 2010, \aap, 518, L2

\bibitem[{{Preibisch} \& {Feigelson}(2005)}]{2005ApJS..160..390P}
{Preibisch}, T. \& {Feigelson}, E.~D. 2005, \apjs, 160, 390

\bibitem[{{Raymond} {et~al.}(2004){Raymond}, {Quinn}, \&
  {Lunine}}]{2004Icar..168....1R}
{Raymond}, S.~N., {Quinn}, T., \& {Lunine}, J.~I. 2004, \icarus, 168, 1

\bibitem[{{Rayner} {et~al.}(2009){Rayner}, {Cushing}, \&
  {Vacca}}]{2009ApJS..185..289R}
{Rayner}, J.~T., {Cushing}, M.~C., \& {Vacca}, W.~D. 2009, \apjs, 185, 289

\bibitem[{{Rieke} {et~al.}(2004)}]{2004ApJS..154...25R}
{Rieke}, G.~H. {et~al.} 2004, \apjs, 154, 25

\bibitem[{{Robitaille} {et~al.}(2007){Robitaille}, {Whitney}, {Indebetouw}, \&
  {Wood}}]{2007ApJS..169..328R}
{Robitaille}, T.~P., {Whitney}, B.~A., {Indebetouw}, R., \& {Wood}, K. 2007,
  \apjs, 169, 328

\bibitem[{{Russell} {et~al.}(2012){Russell}, {Rudy}, {Rossano}, {Kim}, {Laag},
  {Crawford}, {Skinner}, {Gregory}, \& {Sitko}}]{basscalib}
{Russell}, R.~W., {Rudy}, R.~J., {Rossano}, G.~S., {Kim}, D.~L., {Laag}, E.,
  {Crawford}, K., {Skinner}, M.~A., {Gregory}, S.~A., \& {Sitko}, M.~L. 2012,
  in 21st CalCon (Space Dynamics Laboratory)

\bibitem[{{Schegerer} {et~al.}(2008){Schegerer}, {Wolf}, {Ratzka}, \&
  {Leinert}}]{2008A&A...478..779S}
{Schegerer}, A.~A., {Wolf}, S., {Ratzka}, T., \& {Leinert}, C. 2008, \aap, 478,
  779

\bibitem[{{Schmitt} {et~al.}(1995){Schmitt}, {Fleming}, \&
  {Giampapa}}]{1995ApJ...450..392S}
{Schmitt}, J.~H.~M.~M., {Fleming}, T.~A., \& {Giampapa}, M.~S. 1995, \apj, 450,
  392

\bibitem[{{Schneider} {et~al.}(2013){Schneider}, {Song}, {Melis}, {Zuckerman},
  {Bessell}, {Hufford}, \& {Hinkley}}]{2013arXiv1308.0405S}
{Schneider}, A., {Song}, I., {Melis}, C., {Zuckerman}, B., {Bessell}, M.,
  {Hufford}, T., \& {Hinkley}, S. 2013, ArXiv e-prints

\bibitem[{{Sicilia-Aguilar} {et~al.}(2006){Sicilia-Aguilar}, {Hartmann},
  {F{\"u}r{\'e}sz}, {Henning}, {Dullemond}, \&
  {Brandner}}]{2006AJ....132.2135S}
{Sicilia-Aguilar}, A., {Hartmann}, L.~W., {F{\"u}r{\'e}sz}, G., {Henning}, T.,
  {Dullemond}, C., \& {Brandner}, W. 2006, \aj, 132, 2135

\bibitem[{{Sicilia-Aguilar} {et~al.}(2011){Sicilia-Aguilar}, {Henning},
  {Dullemond}, {Patel}, {Juh{\'a}sz}, {Bouwman}, \&
  {Sturm}}]{2011ApJ...742...39S}
{Sicilia-Aguilar}, A., {Henning}, T., {Dullemond}, C.~P., {Patel}, N.,
  {Juh{\'a}sz}, A., {Bouwman}, J., \& {Sturm}, B. 2011, \apj, 742, 39

\bibitem[{{Sicilia-Aguilar} {et~al.}(2010){Sicilia-Aguilar}, {Henning}, \&
  {Hartmann}}]{2010ApJ...710..597S}
{Sicilia-Aguilar}, A., {Henning}, T., \& {Hartmann}, L.~W. 2010, \apj, 710, 597

\bibitem[{{Sicilia-Aguilar} {et~al.}(2013){Sicilia-Aguilar}, {Henning}, {Linz},
  {Andr{\'e}}, {Stutz}, {Eiroa}, \& {White}}]{2013A&A...551A..34S}
{Sicilia-Aguilar}, A., {Henning}, T., {Linz}, H., {Andr{\'e}}, P., {Stutz}, A.,
  {Eiroa}, C., \& {White}, G.~J. 2013, \aap, 551, A34

\bibitem[{{Siess} {et~al.}(2000){Siess}, {Dufour}, \&
  {Forestini}}]{2000A&A...358..593S}
{Siess}, L., {Dufour}, E., \& {Forestini}, M. 2000, \aap, 358, 593

\bibitem[{{Sitko} {et~al.}(2008){Sitko}, {Carpenter}, {Kimes}, {Wilde},
  {Lynch}, {Russell}, {Rudy}, {Mazuk}, {Venturini}, {Puetter}, {Grady},
  {Polomski}, {Wisnewski}, {Brafford}, {Hammel}, \&
  {Perry}}]{2008ApJ...678.1070S}
{Sitko}, M.~L., {Carpenter}, W.~J., {Kimes}, R.~L., {Wilde}, J.~L., {Lynch},
  D.~K., {Russell}, R.~W., {Rudy}, R.~J., {Mazuk}, S.~M., {Venturini}, C.~C.,
  {Puetter}, R.~C., {Grady}, C.~A., {Polomski}, E.~F., {Wisnewski}, J.~P.,
  {Brafford}, S.~M., {Hammel}, H.~B., \& {Perry}, R.~B. 2008, \apj, 678, 1070

\bibitem[{{Skrutskie} {et~al.}(1990){Skrutskie}, {Dutkevitch}, {Strom},
  {Edwards}, {Strom}, \& {Shure}}]{1990AJ.....99.1187S}
{Skrutskie}, M.~F., {Dutkevitch}, D., {Strom}, S.~E., {Edwards}, S., {Strom},
  K.~M., \& {Shure}, M.~A. 1990, \aj, 99, 1187

\bibitem[{{Smith} {et~al.}(2009){Smith}, {Wyatt}, \&
  {Haniff}}]{2009A&A...503..265S}
{Smith}, R., {Wyatt}, M.~C., \& {Haniff}, C.~A. 2009, \aap, 503, 265

\bibitem[{{Smith} {et~al.}(2012){Smith}, {Wyatt}, \&
  {Haniff}}]{2012MNRAS.422.2560S}
---. 2012, \mnras, 422, 2560

\bibitem[{{Soderblom}(2010)}]{2010ARA&A..48..581S}
{Soderblom}, D.~R. 2010, \araa, 48, 581

\bibitem[{{Tetzlaff} {et~al.}(2011){Tetzlaff}, {Neuh{\"a}user}, \&
  {Hohle}}]{2011MNRAS.410..190T}
{Tetzlaff}, N., {Neuh{\"a}user}, R., \& {Hohle}, M.~M. 2011, \mnras, 410, 190

\bibitem[{{Tilling} {et~al.}(2012){Tilling}, {Woitke}, {Meeus}, {Mora},
  {Montesinos}, {Riviere-Marichalar}, {Eiroa}, {Thi}, {Isella}, {Roberge},
  {Martin-Zaidi}, {Kamp}, {Pinte}, {Sandell}, {Vacca}, {M{\'e}nard},
  {Mendigut{\'{\i}}a}, {Duch{\^e}ne}, {Dent}, {Aresu}, {Meijerink}, \&
  {Spaans}}]{2012A&A...538A..20T}
{Tilling}, I., {Woitke}, P., {Meeus}, G., {Mora}, A., {Montesinos}, B.,
  {Riviere-Marichalar}, P., {Eiroa}, C., {Thi}, W.-F., {Isella}, A., {Roberge},
  A., {Martin-Zaidi}, C., {Kamp}, I., {Pinte}, C., {Sandell}, G., {Vacca},
  W.~D., {M{\'e}nard}, F., {Mendigut{\'{\i}}a}, I., {Duch{\^e}ne}, G., {Dent},
  W.~R.~F., {Aresu}, G., {Meijerink}, R., \& {Spaans}, M. 2012, \aap, 538, A20

\bibitem[{{Torres} {et~al.}(2008){Torres}, {Quast}, {Melo}, \&
  {Sterzik}}]{2008hsf2.book..757T}
{Torres}, C.~A.~O., {Quast}, G.~R., {Melo}, C.~H.~F., \& {Sterzik}, M.~F. 2008,
  in Handbook of Star Forming Regions, Volume II, ed. B.~{Reipurth}, 757

\bibitem[{{van Leeuwen}(2007)}]{2007A&A...474..653V}
{van Leeuwen}, F. 2007, \aap, 474, 653

\bibitem[{{Watson} {et~al.}(2009)}]{2009A&A...493..339W}
{Watson}, M.~G. {et~al.} 2009, \aap, 493, 339

\bibitem[{{Weidenschilling}(1977)}]{1977Ap&SS..51..153W}
{Weidenschilling}, S.~J. 1977, \apss, 51, 153

\bibitem[{{Werner} {et~al.}(2004)}]{2004ApJS..154....1W}
{Werner}, M.~W. {et~al.} 2004, \apjs, 154, 1

\bibitem[{{Williams} \& {Cieza}(2011)}]{2011ARA&A..49...67W}
{Williams}, J.~P. \& {Cieza}, L.~A. 2011, \araa, 49, 67

\bibitem[{{Wyatt}(2008)}]{2008ARA&A..46..339W}
{Wyatt}, M.~C. 2008, \araa, 46, 339

\bibitem[{{Wyatt} {et~al.}(2005){Wyatt}, {Greaves}, {Dent}, \&
  {Coulson}}]{2005ApJ...620..492W}
{Wyatt}, M.~C., {Greaves}, J.~S., {Dent}, W.~R.~F., \& {Coulson}, I.~M. 2005,
  \apj, 620, 492

\bibitem[{{Wyatt} {et~al.}(2007){Wyatt}, {Smith}, {Greaves}, {Beichman},
  {Bryden}, \& {Lisse}}]{2007ApJ...658..569W}
{Wyatt}, M.~C., {Smith}, R., {Greaves}, J.~S., {Beichman}, C.~A., {Bryden}, G.,
  \& {Lisse}, C.~M. 2007, \apj, 658, 569

\bibitem[{{Yang} {et~al.}(2012){Yang}, {Herczeg}, {Linsky}, {Brown},
  {Johns-Krull}, {Ingleby}, {Calvet}, {Bergin}, \&
  {Valenti}}]{2012ApJ...744..121Y}
{Yang}, H., {Herczeg}, G.~J., {Linsky}, J.~L., {Brown}, A., {Johns-Krull},
  C.~M., {Ingleby}, L., {Calvet}, N., {Bergin}, E., \& {Valenti}, J.~A. 2012,
  \apj, 744, 121

\bibitem[{{Zuckerman} \& {Becklin}(1993)}]{1993ApJ...406L..25Z}
{Zuckerman}, B. \& {Becklin}, E.~E. 1993, \apjl, 406, L25

\bibitem[{{Zuckerman} {et~al.}(1995){Zuckerman}, {Forveille}, \&
  {Kastner}}]{1995Natur.373..494Z}
{Zuckerman}, B., {Forveille}, T., \& {Kastner}, J.~H. 1995, \nat, 373, 494

\bibitem[{{Zuckerman} \& {Song}(2004)}]{2004ARA&A..42..685Z}
{Zuckerman}, B. \& {Song}, I. 2004, \araa, 42, 685

\end{thebibliography}

\end{document}